\documentstyle[12pt,epsf]{article}
\makeatletter

\@addtoreset{equation}{section}
\makeatother
\begin{document}
\title{Transport coefficients for shape degrees in terms of Cassini
ovaloids
\thanks{Supported in part by the Deutsche Forschungsgemeinschaft}}
\author{F.A.Ivanyuk$^{1,2}$, H.Hofmann
\thanks{e-mail: ~hhofmann@physik.tu-muenchen.de}
\thanks{www home page: ~http://www.physik.tu-muenchen.de/tumphy/e/T36/hofmann.html}
$^{1}$, V.V.Pashkevich$^{3}$  and S.Yamaji$^{4}$  
\\
\small\it{1) Physik-Department der Technischen Universit\"at M\"unchen,
D-85747 Garching, Germany}
\\
\small\it{2) Institute for Nuclear Research of the Ukrainian Academy of
Sciences, Kiev-28, Ukraine}
\\
\small\it{3) Joint Institute for Nuclear Research, 141980 Dubna, Russia}
\\
\small\it{4) Cyclotron Lab., Riken, Wako, Saitama, 351-01, Japan}}
\maketitle
\begin{abstract}
Previous computations of the potential landscape with the shapes
parameterized in terms of Cassini ovaloids are extended to collective
dynamics at finite excitations.  Taking fission as the most demanding
example of large scale collective motion, transport coefficients are
evaluated along a fission path. We concentrate on those for average
motion, namely stiffness $C$, friction $\gamma$ and inertia $M$. Their
expressions are formulated within a locally harmonic approximation and
the help of linear response theory.  Different approximations are
examined and comparisons are made both with previous studies, which
involved different descriptions of single particle dynamics, as well as
with macroscopic models.  Special attention is paid to an appropriate
definition of the deformation of the nuclear density and its relation
to that of the single particle potential. For temperatures above $3 \,
MeV$ the inertia agrees with that of irrotational flow to less than a
factor of two, but shows larger deviations below, in particular in its
dependence on the shape.  Also friction exhibits large fluctuations
along the fission path for small excitations. They get smoothed out
above $3\,- \,4 \, MeV$ where $\gamma$ attains values in the range of
the wall formula.  For $T\ge 2\,MeV$ the inverse relaxation time
$\beta=\gamma /M$ turns out to be rather insensitive to the shape and
increases with $T$.
\end{abstract}
\medskip
\centerline{PACS numbers: 21.60.Ev, 21.60.Cs, 24.10Pa, 24.75+i}
\bigskip
\centerline{\bf to appear in PRC} 

\vskip 1cm

\section{Introduction}

One of the oldest but still most challenging problems of
nuclear physics is an adequate description of collective motion at
finite excitations. As the prime example one may quote
nuclear fission which has attracted the attention of both
experimentalists as well as theoreticians since its discovery. To date
it is still an open question which type of configurations the system
undergoes on its way from the potential minimum over the saddle region
down to scission. Whereas in the early days those of the compound
model were clearly favoured in theoretical pictures, after the
discovery of the shell model that of independent particle motion 
came into fashion more and more. This development was enhanced after
computers got fast enough such that Hartree-Fock type computations
could be done in every lab.

However, there can be little doubt that this picture fails to describe
collective motion at finite excitations where one is compelled almost by
experimental evidence that the dynamics shows irreversible behaviour,
not only by the very nature of the decay process itself but by 
the appearance of frictional forces. It is more than questionable that
this feature can adequately be met by introducing simple minded
collision terms. Decent descriptions of fission in terms of the one
body density operator most likely require to consider correlations
beyond the independent particle picture, together with non-Markovian
effects. This is a difficult problem in itself, not to mention the
computational task of solving this equation of motion for the one body
density. 

For these reasons it may still be interesting and worth while to start
from a more phenomenological point of view introducing the shape
parameters as collective variables. It is true that in this way again
the
picture of independent particles will serve as a starting point, in the
form of the deformed shell model. However, the latter is simple and
flexible enough to allow one considering residual interactions, in one
way or other. As we shall see, it may be possible to gain insight into
their importance by studying dynamical aspects. Likewise, we may be able
to get information on the complexity of the configurations which are to
be considered.
Such a task becomes more feasible in case collective motion is
sufficiently slow. Then one may exploit the quasi-static picture which
reduces the complexity of the full problem drastically. Under such
circumstances one may actually linearize the problem and treat
collective motion locally within a harmonic approximation. 
In this way one may take advantage of the benefits of linear response
theory.  

One of the major problems in theories of this type is to find a decent
guess for the relevant macroscopic variables, a problem which is
familiar almost from all transport theories. For nuclear
fission there exists some kind of guiding principle through the liquid
drop model. The latter is known to represent the static energy for
temperatures above $1\,-\,2\; MeV$. Since at these temperatures one
expects motion to be strongly damped, it will most likely follow
somehow the line of steepest decent. Possible shapes which a fissioning
nucleus may assume on its way to scission have been looked for in 
\cite{stlapo} by minimizing the liquid drop energy. This minimization
has
been done for some realistic energy density functional under the
constraint of fixing a parameter which measures the distance between
the evolving fragments. Incidentally, it is the same parameter which we
are going to exploit later on in our approach. It so turns out that the
shapes found in this way can be approximated fairly 
well by the Cassini ovaloids introduced to nuclear physics in
\cite{strase}.
Later in \cite{pash71} a single particle model has been constructed for
such a parameterization of shapes, which was based on the Woods-Saxon
potential. 

In this paper we are going to use this model for computations of
transport coefficients, after some suitable modifications which are
necessary to incorporate the effects mentioned above. One of our goals
will be to study average motion along the fission path for different
temperatures, as it is reflected in the associated transport
coefficients of inertia, friction and local stiffness. In this sense the
aim of our present work is similar to the one of \cite{yaivho}, where a 
two center shell model was used.  The latter feature renders the
previous model simpler on the computational level. On the other hand,
the parameterization of the  shape by means of Cassini ovaloids opens
the possibility of treating more realistic shapes, which are perhaps
better suited to describe the later stages of a fission process.
Furthermore, it is fair to say that the Woods-Saxon potential may be
supposed to resemble more the "true" mean field.

For Cassini ovaloids commonly a few parameters suffice to treat in  
simple terms a whole variety of realistic shapes including very compact
ones as well as strongly deformed ones with a well developed neck, or
even those corresponding to separated fragments. In this sense this
parameterization may be considered superior to expansions
in terms of spherical harmonics (see \cite{bohrmo2}). In
the ideal case one would then be able to compute transport tensors
for all the parameters, the collective degrees of freedom, one claims
to be relevant. This is a tremendous task and so far has been carried
through only for a two dimensional model \cite{samhahy}, without
utilizing though the full microscopic potential of linear response
theory. In this paper we want to restrict ourselves to the one
dimensional case. The main reason for that is found, of course, in the
simplification one gains by this restriction. However, it may be said
that at present most of the applications of macroscopic equations of
motion to fission at finite excitation adhere to a similar confinement,
see e.g. \cite{wada} - \cite{thoenn}. Evidently one then needs to rely
on the "right" guess of the fission path. As said before and for
arguments given there we presume it to be represented well enough by
the line along the valley of the static energy. Possible improvements
have to be left for future studies.

\section{ Deformed shell model}

\subsection{ Shape parameterization}

We follow the suggestion put forward in \cite{pash71}, but would like to
repeat the most important elements for convenience.
The Cassini ovaloids are obtained by rotating the curve
\begin{equation}
\label{ovals}
\rho(z, \epsilon)=R_0\left[\sqrt{a^4+4\epsilon z^2/R_0^2}-z^2/R_0^2- 
     \epsilon\right]^{1/2}    
\end{equation}
around the $z$-axis, with $z$ and $\rho$ being cylindrical coordinates. 
The constant $a$ is defined by volume conservation, implying that the
family of shapes (\ref{ovals}) depends only on one deformation parameter 
$\epsilon$. As is easily recognized from (\ref{ovals}) the value of
$\epsilon =0$ corresponds to a sphere. For $0 < \epsilon < 0.4$ 
the form resembles very much that of a spheroid with the ratio of
the axes given by
\begin{equation}
\label{ratio}
{\rm{shorter\,\,\,axes}\over \rm{longer\,\,\,axes}}=
    {{1-2\epsilon /3}\over{1+\epsilon /3}} 
\end{equation}
At $\epsilon \approx 0.5$ a neck appears and at $\epsilon =1.0$ the 
nucleus
separates into two fragments. A few examples of the family (\ref{ovals})
are shown in  Fig.1.  

It is possible to describe a more general class of axially symmetric
shapes by exploiting an expansion about the surface given by
(\ref{ovals}). We may introduce two new coordinates $R$ and $x$ such
that $R=const=R_0$ corresponds to the Cassini ovals (\ref{ovals}).
The coordinate $x$ specifies the position of a point on
the line given by (\ref{ovals}), see \cite{pash71} for details. With
these two variables at our disposal we may parameterize a new shape.
The latter is meant to  express the  deviation from the ovaloid given by
(\ref{ovals}) by means of an expansion into a series of Legendre
polynomials
\begin{equation}
\label{alphas}
R(x)=R_0(1+\sum_n\alpha_n P_n(x)) 
\end{equation}
The full set of collective variables or parameters then includes
the coefficients $\alpha_n$ in addition to the $\epsilon $  from
before.  
\begin{figure}[hb]
\centerline{{
\epsfysize=8cm
\leavevmode
\epsffile[85 295 408 510]{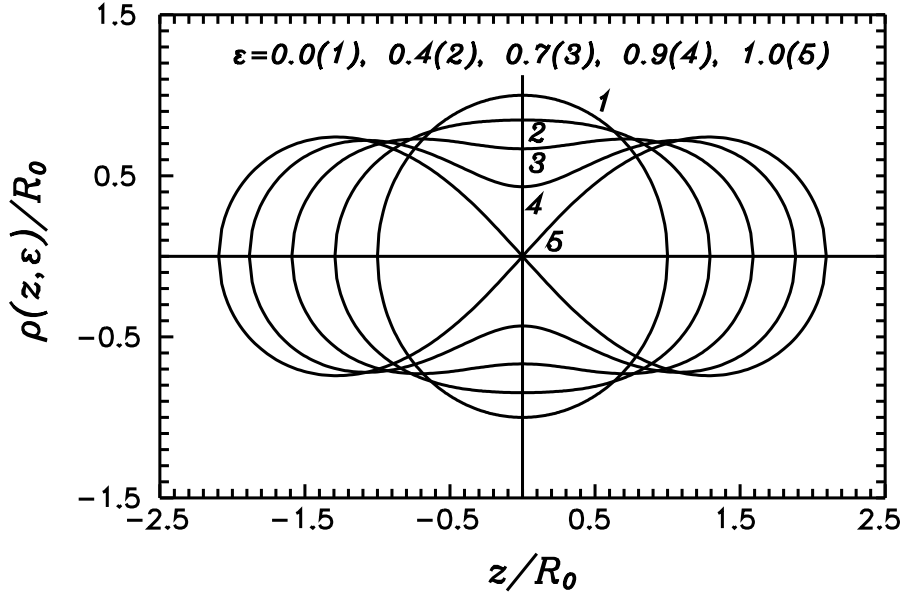}
}}
\caption{The parameterization of the shape of the nuclear surface in
terms of Cassini ovaloids. The values of the deformation parameter
$\epsilon$ (see (\protect\ref{ovals})) are indicated in the Figure. 
}
\label{Fig.1}
\end{figure}
\begin{figure}[hb]
\centerline{{
\epsfysize=7cm
\leavevmode
\epsffile[102 286 502 547]{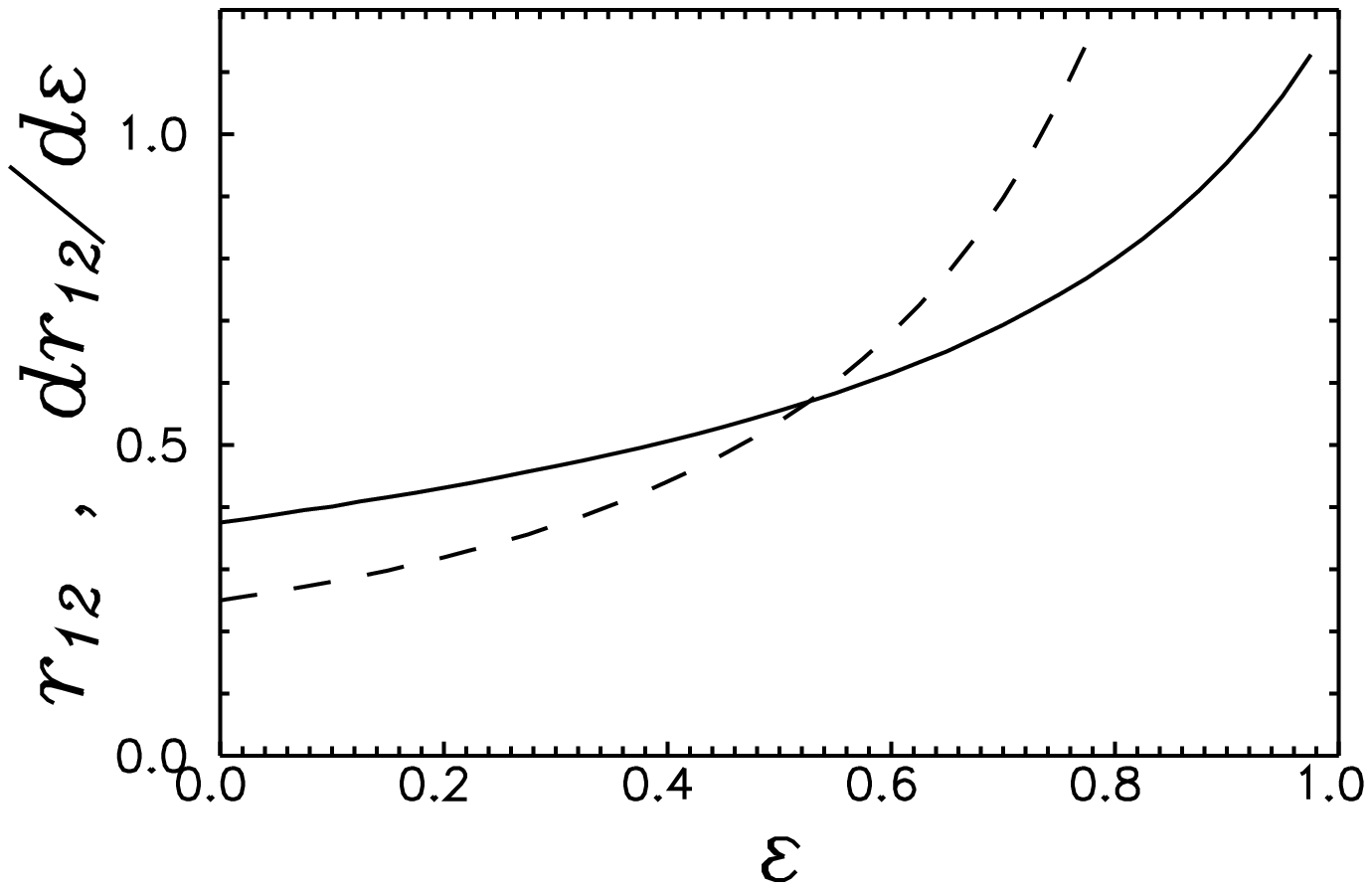}
}}
\caption{The relation between ${r_{12}}$ and  ${\epsilon}$
(solid line); the derivative $dr_{12}/d\epsilon$ is shown by dashed
line.
}
\label{Fig.2}
\end{figure}

Sometimes it is convenient to introduce a measure for the overall
elongation of the nucleus instead of $\epsilon $. One may chose, for
instance, the distance $R_{12}$ between the left and right center of
masses. To have a dimensionless quantity one may divide  $R_{12}$ by the
diameter $2R_0$ of the sphere (of identical volume) to get
\begin{equation}
\label{rr12}
r_{12} \equiv {R_{12}\over 2R_0}
    ={{\int \vert z-z_{cm} \vert dV}
    \over{R_0\int dV}}  
\end{equation}
with $z_{cm}$ being the $z$-coordinate of the center of mass of the
the whole complex. The integration is carried out over the volume 
within the sharp surface specified by (\ref{alphas}).  Asymptotically, 
the $r_{12}$ turns into half of the distance between centers of mass of 
the fission fragments. 
Incidentally, this variable is defined uniquely for any parameterization
of the shape and has been used in the past by many authors. 
This feature facilitates comparisons to theories which
are based on shell models with different shape variables. The relation
between $r_{12}$ and $\epsilon$ is demonstrated in  Fig.2.

\subsection{ Single-particle Hamiltonian}

The single-particle Hamiltonian $\hat h_{ipm}$ will be constructed like
in \cite{pash71}. It has terms for the kinetic energy $\hat T$, the
radial potential $\hat V$, the spin-orbit coupling $\hat V_{s.o.}$ and
the Coulomb potential $\hat V_{Coul}$:
\begin{equation}
\label{hipm}
\hat h_{ipm}=\hat T+\hat V+\hat V_{s.o.}+\hat V_{Coul}                 
\end{equation}
The radial part $\hat V$ is represented by a finite depth Woods-Saxon
potential
\begin{equation}
\label{wspot}
V(\rho, z)=V_0[1+\exp (l(\rho, z)/a)]^{-1}   
\end{equation}
where $l(\rho, z)$ is the shortest distance from the point $(\rho , z)$
to the sharp surface and $a$ is the diffuseness parameter which is
assumed to be constant along the surface.  
The spin-orbit potential may
be written in a way which makes apparent that it is proportional to the
gradient of the potential given in (\ref{wspot}),
\begin{equation}
\label{sopot}
\hat V_{s.o.}\propto [\vec s, \vec p]\nabla V     
\end{equation}
Here $\vec p$ and $\vec s$ stand for the nucleon's momentum and the
spin.
The Coulomb potential is calculated for a charge
distributed uniformly inside the sharp surface (\ref{alphas}) or 
(\ref{ovals}).

The single-particle energies and wave functions are determined by
diagonalizing the matrix of the Hamiltonian (\ref{hipm}) calculated
with the wave functions of a deformed axially symmetric oscillator 
potential, see \cite{dapapa}. An example of the deformation dependence
of the single-particle energies is shown in Fig.3. 
As the result of diagonalization one obtains not only energies and wave
functions of bound states but also those of discrete states of positive
energy, which for the Woods-Saxon potential lie in the continuum. The
density of these states depends on the number of oscillator shells
included into the basis. In the computations within the shell correction
method the number of oscillator shells is optimized by the requirement
that the states with positive energies provide a smooth extrapolation of
the density of bound states into the continuum. Accounting for such
states with positive energy improves considerably the "plateau" of the
shell correction as a function of the averaging interval.  In the
present paper we do so not only when calculating deformation energies
but also in the computation of transport coefficients.  The cut-off in
the single-particle energy was set equal to $20 MeV$. We have checked
that a variation of the cut-off energy within the interval $5-20 MeV$
does not change much the values of the transport coefficients.
This may be understood from the fact that these transport 
coefficients reflect the truly low frequency behavior of the system.
\begin{figure}[ht]
\centerline{{
\epsfysize=7cm
\leavevmode
\epsffile[74 286 465 553]{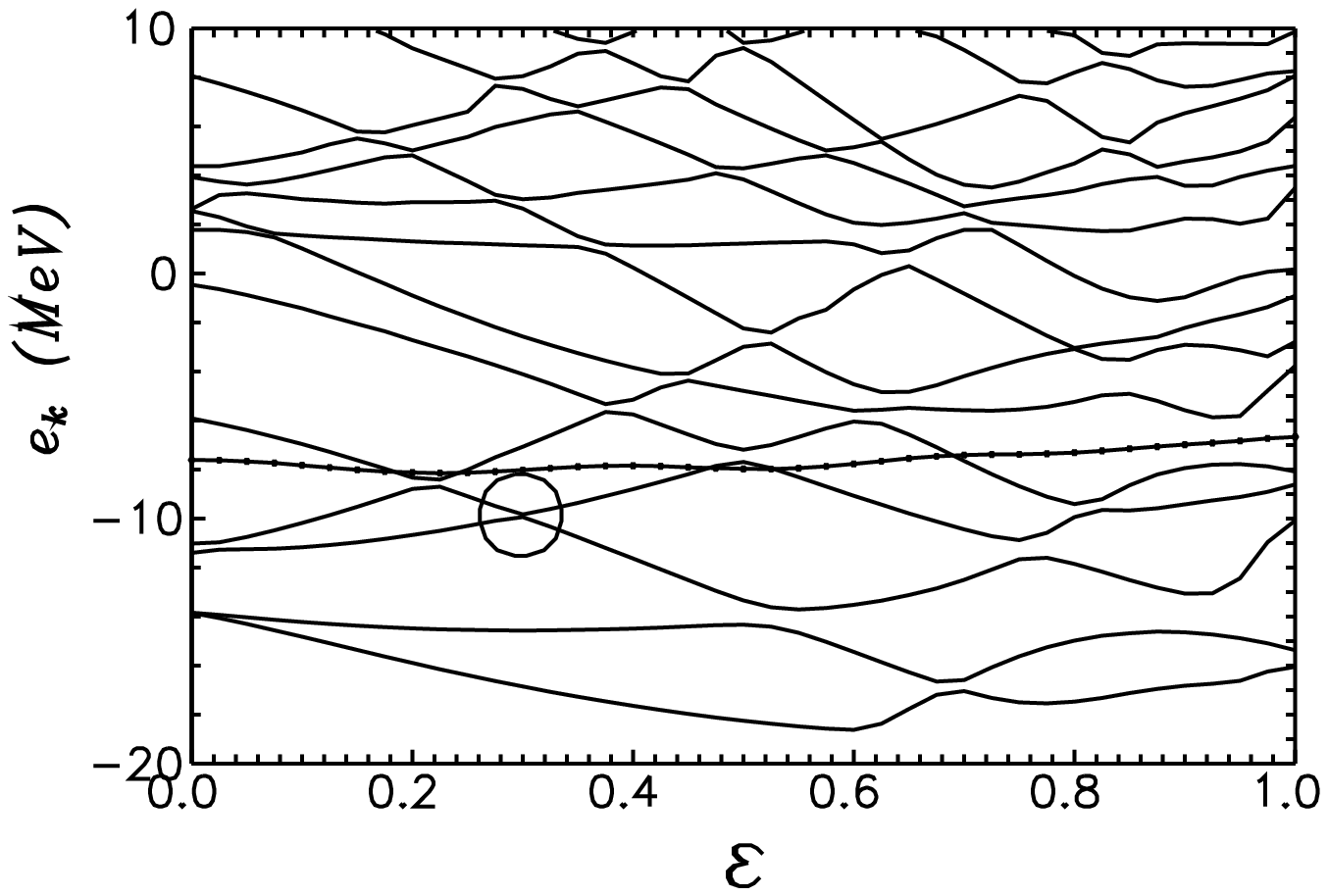}
}}
\caption{The energies $e_k$ of single-particle states
(for fixed $z$-component of angular momentum and parity,
$j_z^{\pi}=3/2^{-}$) as functions of $\epsilon$. The line with stars 
marks
the position of the chemical potential computed for $T=1 MeV$. The
circle marks the pseudo crossing which at $\epsilon=0.3$ is closest to
the chemical potential, see text.
}
\label{Fig.3}
\end{figure} 

\subsection{ Deformation energy}

The deformation energy $E_{def}$ at zero temperature is calculated
according to the shell correction method \cite{strut},\cite{funnyh} as
the sum of the liquid drop energy $E_{def}^{LDM}$ and the shell
correction $\delta E^{n,p}+\delta P^{n,p}$ (including the one for the
pairing energy)
\begin{equation}
\label{enedef}
E_{def}=E_{def}^{LDM}+\sum_{p,n}(\delta E^{p,n}+\delta P^{p,n})
\end{equation}
The liquid drop energy is computed as the
contributions from the Coulomb $E_{Coul}$ and surface $E_{S}$ energies
according to \cite{strut},\cite{funnyh} 
\begin{equation}
\label{wldm}
E_{def}^{LDM}=E_{Coul}+E_{S}-(E_{Coul}^0+E_{S}^0) 
\end{equation}
where $E_{Coul}^0$ and $E_S^0$ are the corresponding energies of the
spherical shape.
As an example,  Fig.4  exhibits the results of the calculation of
$E_{def}$ and $E_{def}^{LDM}$ at zero temperature for $^{224}Th$ as 
function 
of parameters $\epsilon$ and $\alpha_3$.

The temperature dependence of Coulomb and surface energy is accounted
for by using the forms
\begin{equation}
\label{eldmt}
E_{Coul}(T)=E_{Coul}(T=0)(1-\alpha T^2),\,\,\,\,
     E_{S}(T)=E_{S}(T=0)(1-\beta T^2)           
\end{equation}
with $\alpha = 0.000763 MeV^{-2}$ and $\beta =$ $0.00553$ $MeV^{-2}$ 
\cite{gustbr}. To compute the shell correction at finite temperature we
use the phenomenological ansatz proposed in \cite{bohrmo2}
\begin{equation}
\label{tempshco}
\delta E^{p,n}(T)+\delta P^{p,n}(T)=
    \left[\delta E^{p,n}(T=0)+\delta P^{p,n}(T=0)\right]
    {\tau \over \sinh \tau}    
\end{equation}
with $\tau=2\pi^2T/\hbar \omega_0$ and $\hbar \omega_0=41A^{-1/3}$, $A$
\begin{figure}[ht]
\centerline{{
\epsfysize=12cm
\leavevmode
\epsffile[51 173 490 624]{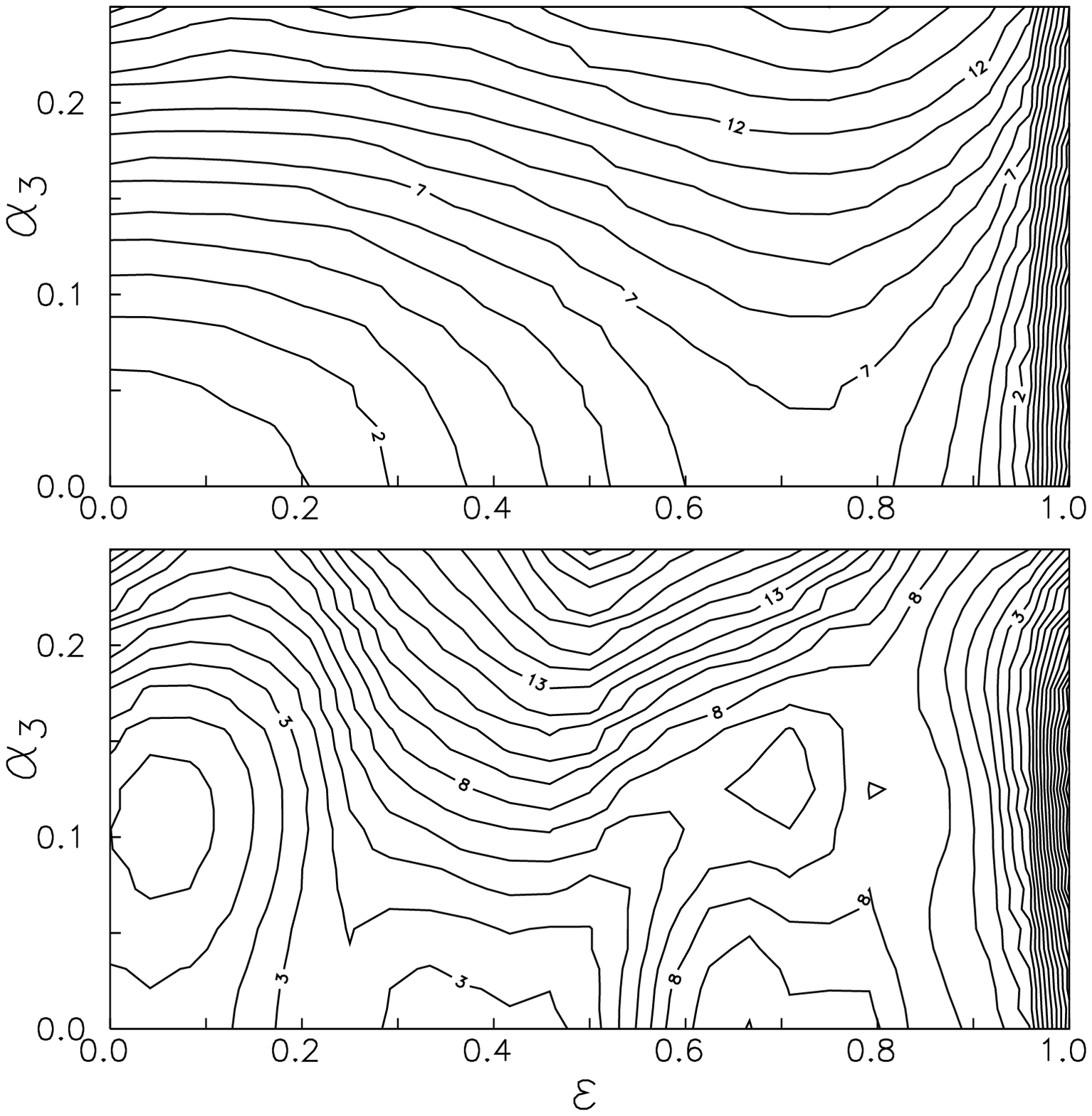}
}}
\caption{The liquid drop (top) and total (bottom) deformation energies 
of ${^{224}Th}$ at $T=0$ as function of $\epsilon$ and  $\alpha_3$. 
}
\label{Fig.4}
\end{figure}
For temperatures larger than $1 MeV$ (as considered in the present
paper)
the shell effects are strongly  suppressed. Already at $T=1 MeV$
the minimum of the total deformation energy almost coincides with the
bottom of the liquid drop valley. As said before the latter can be
approximated rather well just by the Cassini ovaloids.  Thus we
restricted our set of deformation parameters 
to the one parameter $\epsilon$ only, with all the $\alpha_n$ 
put equal to zero. However instead of $\epsilon$ we
prefer to use  $r_{12}$ defined by \ref{rr12}.
The $r_{12}$-dependence of the total
deformation energy and that of the liquid drop is shown in Fig.5.
being the mass number of the fissioning nucleus.

\section{ Dynamics in locally harmonic approximation}

In the following we will assume to be given a many body Hamiltonian
$\hat H(\hat x_i, \hat p_i, Q(t))$ which depends parametrically on the
collective variable $Q$ which specifies the shape of nuclear surface.
Although for the computations to be presented below, it will mostly be
identical to $r_{12}$, in this section we still prefer to use the
general notation $Q$ instead, last but not least to indicate the
general validity of the discussion to come. This Hamiltonian is assumed
to represent the system's total energy. On the level of the shell model
this means to add some c-number terms to the sum over those single
particle Hamiltonians introduced in (\ref{hipm}) (see\cite{sije}).
As will be discussed in the next section, later on we want to account
for collisional damping, which from a principal point of view requires
to add a two body interaction $\hat V^{(2)}_{res}(\hat x_i,
\hat p_i)$. For the moment it is not very important to know details
about the way it will be handled --- besides the fact that we claim
this interaction to be independent of $Q$. 
\begin{figure}[hb]
\centerline{{
\epsfysize=8cm
\leavevmode
\epsffile[102 295 476 564]{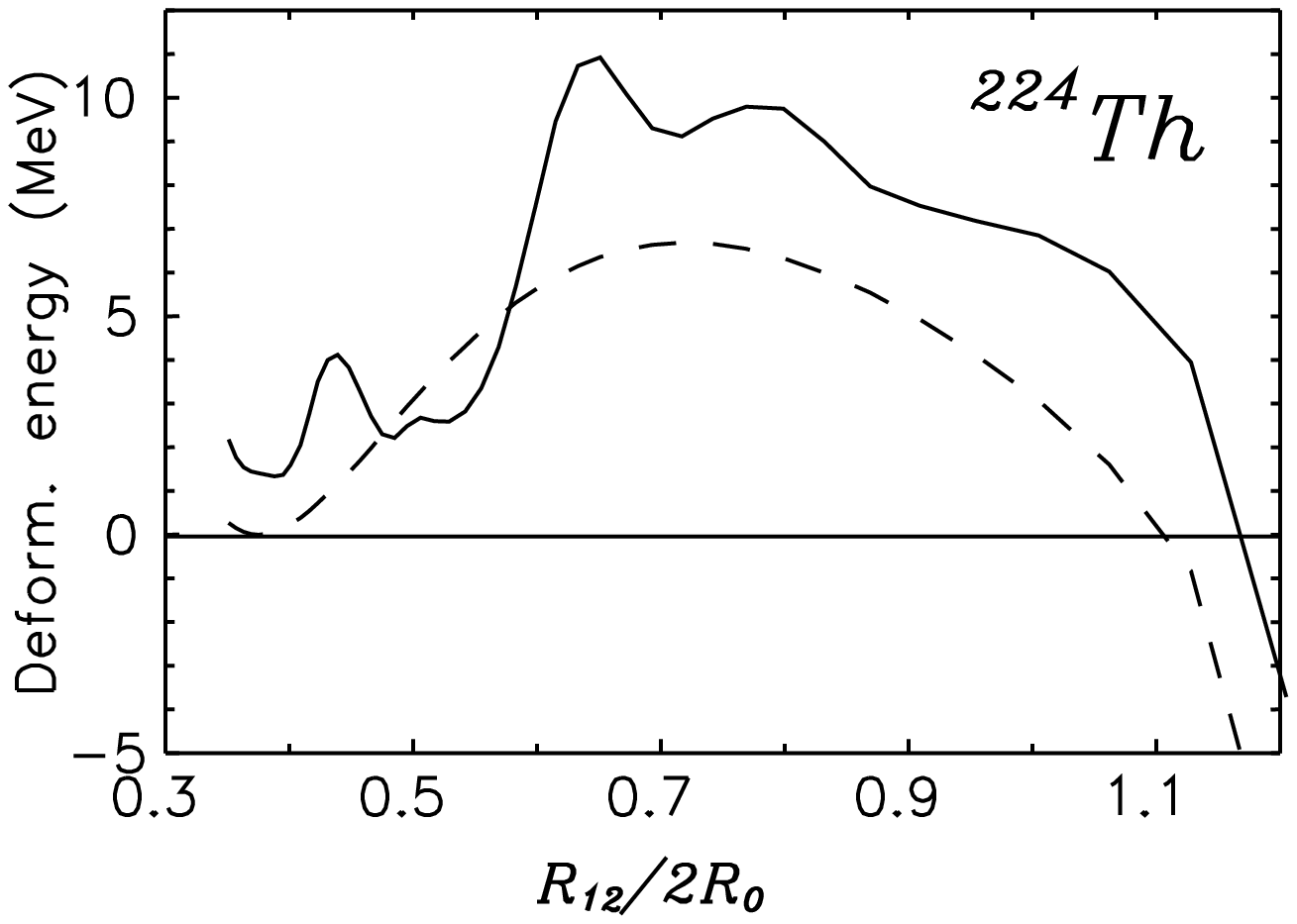}
}}
\caption{The total (solid) and liquid (dashed) drop components of the
deformation  energy along the liquid drop fission valley for  
${^{224}Th}$.
}
\label{Fig.5}
\end{figure}

As a consequence of the latter feature, the generator for collective
motion, namely $\partial \hat H({\hat x}_i,{\hat p}_i,Q) /$ $\partial Q
\equiv  \hat F({\hat x}_i, {\hat p}_i,Q) $, is of pure one body nature.
This operator defines the main source of the coupling between the
collective degree of freedom $Q(t)$ and the nucleonic ones. Indeed,
within the locally harmonic approximation (LHA) the effective
Hamiltonian can be written as
\begin{equation}
\label{hamilapp}
\hat H(Q)=\hat H(Q_0)  + 
     (Q-Q_0)\hat F +     {1\over 2}(Q-Q_0)^2 
     \left<{\partial^2\hat H\over \partial Q^2}(Q_0)
     \right>^{\rm qs}_{Q_0,T_0}  
\end{equation} 
In the second order term the "nucleonic" part appears only
as an average of the corresponding operator. Consistently with the
harmonic approximation, this average is to be built with that density
operator $\rho_{{\rm qs}}(Q_0)$ which in the quasi-static picture is to
be calculated with the Hamiltonian at $Q_0$, namely $\hat H(Q_0)$. It
is here where thermal concepts come into play. In this "unperturbed"
density operator (for the nucleons) one needs to specify the amount of
heat the nucleonic (or intrinsic) degrees of freedom have at the given
configuration parameterized by $Q_0$. The simplest possibility is
offered
by the canonical distribution 
\footnote{For a discussion of the general problems of using
the concept of temperature for an isolated system see \cite{book}.
}
$\rho_{{\rm qs}}(Q_0) \propto  \exp(-\hat H(Q_0)/T)$, to which our
computations will be restricted. Clearly this picture has to rely on
the assumption of a quick relaxation of the relevant internal degrees
of freedom; we will come back to this question later on.

Details about this LHA can be found in many references, see e.g.
\cite{book},\cite{kidhofiva}, \cite{hoivya}: There it is also described 
how this
local dynamics can be handled within a suitable application of linear
response theory. For this reason, we will only recall the most
important theoretical issues.

\subsection{ Collective response function}

The local motion in the $Q$-variable can be described in terms of the so
called collective response function $\chi_{\rm coll}(\omega)$. It can be
derived by introducing a (hypothetical) external force $\hat F\tilde
f_{ext}(t)$ and by evaluating how the deviation of $\big\langle \hat F
\big\rangle_{\omega}$ from some properly chosen static value reacts to
this external field in linear order:
\begin{equation}
\label{eom}
\delta \big\langle \hat F \big\rangle_{\omega}=-\chi_{\rm coll(\omega)} 
f_{ext}(\omega) 
\end{equation}
As shown in \cite{kidhofiva} and \cite{book} the $\chi_{\rm
coll}(\omega)$ can be brought to the form
\begin{equation}
\label{rfcoll}
\chi_{\rm coll(\omega)} = {\chi(\omega)\over 1+k\chi(\omega)}
\end{equation}
Here a response function $\chi(\omega)$  for "intrinsic" motion appears: 
The
$\chi(\omega)$ measures how, at some given shape $Q_0$ and for some
temperature $T_0$, the nucleonic degrees of freedom react to the
coupling $\hat F \delta Q(\omega)$. Its time dependent version reads
\begin{equation}
\label{tdepresp}
\tilde{\chi}(t-s)=\Theta(t-s)\,{i\over\hbar}\,
    {\rm tr}(\hat{\rho}_{qs}(Q_{0},T_{0})[\hat{F}(t),\hat{F}(s)])
    \equiv 2i\Theta(t-s)\tilde{\chi}^{\prime\prime}(t-s)
\end{equation}
In this expression the time development of the field operators is
defined by the same Hamiltonian $\hat H({\hat x}_i,{\hat p}_i,Q_0)$
which appears in the density $\hat \rho_{{\rm qs}}$. The function
$\tilde{\chi}^{\prime\prime}(t-s)$ on the very right stands for the so
called dissipative part. In Fourier space the full response separates
into real and imaginary parts like $\chi(\omega)=\chi^\prime(\omega) + i
\chi^{\prime \prime}(\omega)$, the $\chi^\prime(\omega)$ sometimes
being called reactive part. 

As one may guess from the very construction, the derivation of
(\ref{rfcoll}) relies on quasi-static properties of the nucleonic
degrees
of freedom. For instance, such properties appear in the coupling
constant  $k$ which is to be determined from 
\begin{equation}
\label{appa}
-k^{-1}=\left({\partial^2E\over\partial Q^2}\right)_S
    +\chi(\omega=0) \equiv C(0)+\chi(0)    
\end{equation}
Moreover, the nucleonic degrees of freedom are assumed to behave
ergodic in the sense of having the adiabatic susceptibility
$ \chi^{\rm ad} = -\delta \big\langle \hat F \big\rangle/\delta
Q_{S}$ be identical to the isolated one, the static response
$\chi(0) $:
\begin{equation}
\label{ergodic}
\chi(0) = \chi^{\rm ad} 
\end{equation}
As has been demonstrated in \cite{hoivya} this condition is
not fulfilled in the deformed shell model, which implies that
special measures are to be taken to which we will come to below.

It should be noted that the derivation of (\ref{rfcoll}) involves a
self-consistency relation between the deformation $Q$ of the mean field
and the one of the density. The latter may be measured by the
expectation value $\big\langle \hat F \big\rangle_{t}$. For
linearized dynamics this self-consistency condition reduces to  the
equation 
\begin{equation}
\label{selfcond}
k \, \big\langle \hat F \big\rangle_{t}
     = Q(t) - Q_0 
\end{equation}
well known from the case of undamped vibrations \cite{bohrmo2},
\cite{sije}.  Realize, please, that the quantity $\big\langle \hat F
\big\rangle_{t}$  is to be calculated with the actual dynamical nuclear
states accounting for their appropriate occupations.

Before concluding this subsection we like to write down a more
convenient form for the stiffness $C(0)$ appearing in (\ref{appa}). It
is
defined as the second derivative of the internal energy $E(Q,S)$ with
respect to deformation at fixed entropy $S$.  Since it is not easy to
calculate such a derivative it is
better to reexpress it by the one of the free energy $f$ at fixed
temperature (see eqs.(A.18)-(A.19) of \cite{kidhofiva})
\begin{equation}
\label{stiff0}
C(0)= \left({\partial^2E\over \partial Q^2}\right)_S
     = \left({{\partial^2f}\over{\partial Q^2}}\right)_T
     +\left({\partial T\over \partial Q}\right)_S
    {{\partial^2f(Q,T)}\over{\partial Q \partial T}}    
\end{equation}
In \cite{kidhofiva} it was found that for temperatures larger than
$1.5-2.0 MeV$ the change of $T$ with the collective coordinate $Q$ is
small such that the second term on the right hand side of 
eq.(\ref{stiff0})
can be neglected. Below we would like to use the shell correction
method when calculating static energies. In this method usually the
intrinsic energy is involved, rather than the free energy. Thus one
needs to relate the derivatives of the free and the intrinsic energy
both taken at at fixed temperature. This can be done by differentiating
the relation $E=f+TS$ with respect to deformation and obeying that the
entropy can be expressed as $S = -(\partial f / \partial T)_Q$. As the
result one gets
\begin{equation}
\label{stiffe}
C(0)\approx \left({\partial^2f\over \partial Q^2}\right)_T
     = \left({\partial^2E \over\partial Q^2}\right)_T
     +T{\partial \over \partial T}  
     \left({\partial^2f\over \partial Q^2}\right)_T
    \approx \left({\partial^2E\over\partial Q^2}\right)_T   
\end{equation}
Like before, the expression on the very right is justified for larger
temperatures, where the change of $\partial^2 f/\partial Q^2$
with $T$ is small. This approximation is used in the
computations presented below.  In  Fig.6 the stiffness (\ref{stiffe}) of
the energy (\ref{enedef})-(\ref{tempshco}) is shown as function of the
deformation parameter $r_{12}$.  It is seen that its liquid drop part
becomes negative for $r_{12} \geq 0.5$  and the total stiffness exhibits
rather strong fluctuations due to shell structure.
\begin{figure}[h]
\centerline{{
\epsfysize=8cm
\leavevmode
\epsffile[96 289 522 547]{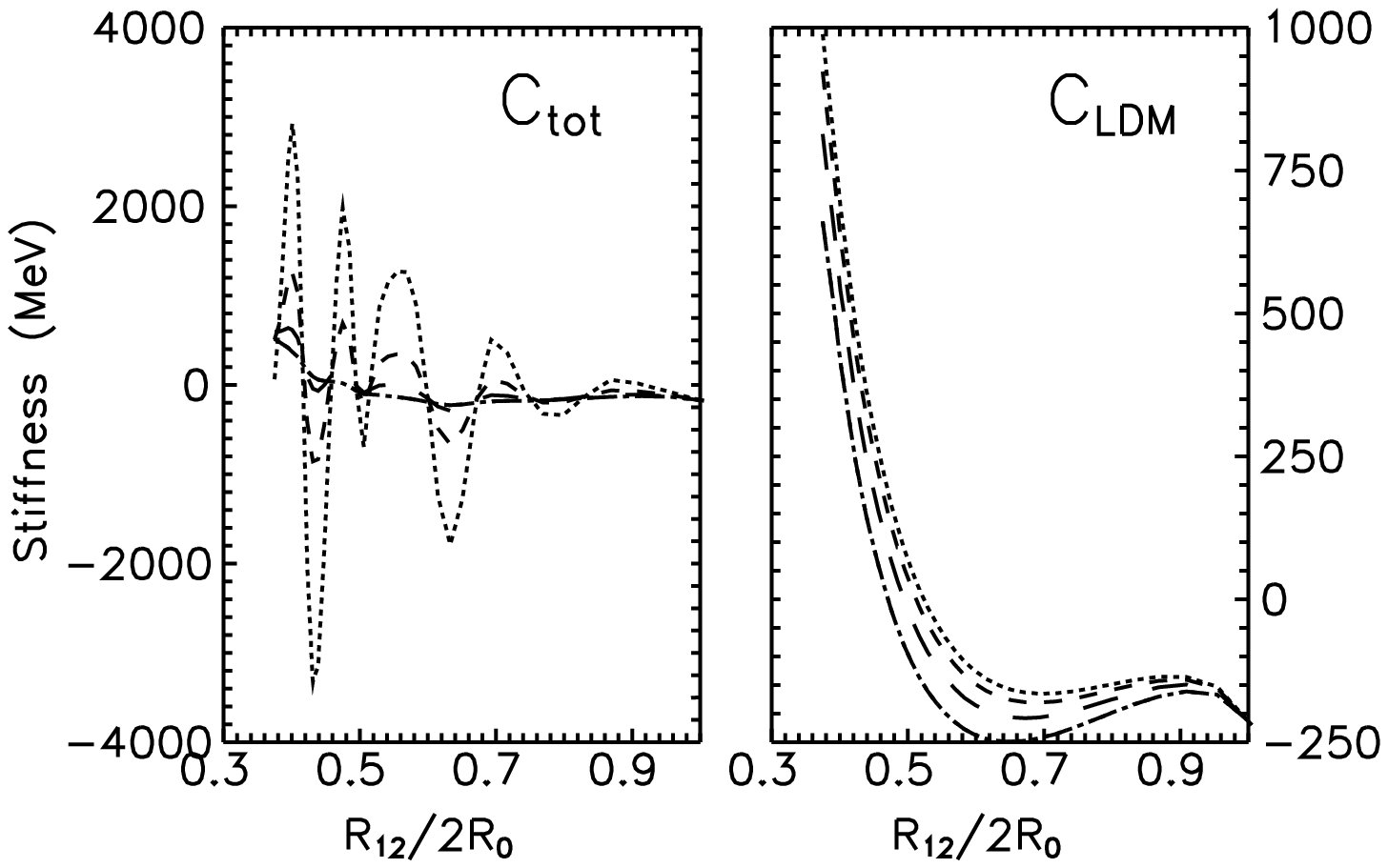}
}}
\caption{The zero frequency limit (\protect\ref{stiff0}) for the
stiffness 
of the total (left) and liquid drop (right) static energies.  The
dotted, short-dashed, long-dashed and dotted-dashed lines correspond to
temperatures $T=1,2,3$ and $4 MeV$.
}
\label{Fig.6}
\end{figure}

\subsection{ Transport coefficients }

In general the frequency dependence of $\chi_{\rm coll}(\omega)$
exhibits a complex structure. Its dissipative part $\chi^{{\prime
\prime}}_{\rm coll}(\omega)$ represents the strength distribution over
all possible modes of the whole nucleus which can be excited by an
external force like the one introduced above, namely $\hat F\tilde
f_{ext}(t)$. Rewriting (\ref{eom}) as
\begin{equation}
\label{nonmarkeom}
\left(\chi_{\rm coll(\omega)}\right) ^{-1}\; 
    \delta \langle \hat F \rangle_\omega  \
    = -  f_{\rm ext}(\omega)    
\end{equation}
it follows that the inverse of the collective response function
can be interpreted as an integral kernel for the effective equation of
motion for the time dependent quantity  $\delta \langle \hat F 
\rangle_t$. Evidently, this time dependent form of the equation of
motion must be expected to contain non-Markovian effects.  However,
there may be situations for which it becomes possible to reduce this
complicated structure to differential form. The clue to this
simplification can be found by recalling the case of the damped
oscillator for which (\ref{nonmarkeom}) takes on the form
\begin{equation}
\label{eomosc}
\left(\chi_{\rm osc}(\omega) \right) ^{-1}\; 
    \delta \langle \hat F \rangle_{\omega}  =
    \left(-M^F\omega^2 -\gamma^F i \omega + C^F\right)\; 
    \delta \langle \hat F \rangle_{\omega} 
    = -  f_{\rm ext}(\omega)    
\end{equation}
Here the following effective "forces" appear: an inertial one, a
friction
force and the conservative one which is related to the derivative of
the effective collective potential, calculated in linearized form. The
associated "transport coefficients" have been marked by an upper index
"F", to indicate that they are associated to the quantity $\delta
\langle \hat F \rangle$. 

Now the strategy of how to handle the general case is clear: Whenever
there is a pronounced peak in the strength distribution $\chi^{{\prime
\prime}}_{\rm coll}(\omega)$ we may approximate it by a Lorentzian.
Since the latter is defined by three quantities, like the width, the
position of the maximum and its height, one may deduce the three
transport coefficients which appear in the oscillator response, for
instance as given by (\ref{eomosc}). In principle, such a procedure may
be applied to any one of the peaks of the original strength
distribution, which allows one to deduce these transport coefficients
for all the possible collective modes, the low frequency ones as well as
the high frequency ones. As a matter of fact, such a scheme is commonly
adopted for the description of collective vibrations, as discussed in
\cite{bohrmo2}, \cite{sije}. The difference to the present application
is seen in the fact that we want to apply this procedure to describe
global dynamics in locally harmonic fashion. The latter aspect puts an
additional constraint.  As mentioned earlier, our application of linear
response theory goes along with the assumption that collective motion is
slow, such that the nucleonic degrees of freedom follow closely a
thermal equilibrium. We may recall that in some formulas given above,
this equilibrium has been parameterized by the quasi-static density
operator $\hat\rho_{{\rm qs}}(Q,T)$.  Estimates of the time scale on
which such a relaxation can be expected to occur will be given below.
But already on this level of information, it is clear that the whole
concept would probably not work if high frequency collective modes would
be important. For this reason we are bound to concentrate on low
frequency ones, if the construction of the transport coefficients is to
be consistent with the basic assumptions for the applicability of the
quasi-static picture.

In practice the "fit" of the strength distributions of the oscillator
model to the "correct" one involves the full response functions, not
only their dissipative parts. In short such an adjustment may be
characterized by $ \left(\chi_{\rm coll}(\omega)\right) ^{-1}\; \delta
\langle \hat F \rangle_\omega  \simeq \left(\chi_{\rm osc}(\omega)
\right) ^{-1}\; \delta \langle \hat F \rangle_\omega $. A more correct
form can be written as a variational procedure
\begin{equation}
\label{lorfit}
\delta \int_0^{\omega_{max}}\Bigl\arrowvert 
    k^2\chi_{\rm coll}(\omega) -{1\over {-M(\omega_1)}\omega^2
    -i\gamma (\omega_1)\omega
    +C(\omega_1)}\Bigr\arrowvert^2 d \omega = 0    
\end{equation}
with the variation to be performed with respect to the coefficients
$C(\omega_1),~\gamma (\omega_1)$ and $M(\omega_1)$.  In this way both
real and imaginary parts of $\chi_{\rm coll}(\omega)$ are fitted
simultaneously. Fortunately, for the practical applications to be
discussed below it turns out that the coefficients $C(\omega_1), \gamma
(\omega_1)$ and $M(\omega_1)$ are rather insensitive to the upper
integration limit $\omega_{max}$. The value of the latter was fixed to
be $\hbar \omega_{max} =5MeV$.

The reader will have noticed the appearance of the factor $k^2$ in
(\ref{lorfit}), as well as the fact that we have left out the upper
index $F$ in the transport coefficients. This change is easily
understood by referring to the self-consistency condition
(\ref{selfcond}). The latter can be interpreted as a transformation from
the $F$-mode to the $Q$-mode. It implies a corresponding transformation
both for the response functions as well as for the transport
coefficients (for details see \cite{book}). 
In this sense the  $k^2\chi_{\rm coll}(\omega)$ and the transport
coefficients ${\cal T}=M,\gamma,C$  stand for the collective response
and the coefficients of the $Q$-mode, respectively, with ${\cal
T}^F=k^2{\cal T}$.

It may turn out that collective motion is so slow, as compared to the
dynamics of the nucleons, that the transport coefficients can be deduced
by expanding the response functions around $\omega=0$. At first such a
procedure has been applied to the intrinsic response (cf.\cite{holet}).
In this way one may get an approximate solution of the secular equation
for the position of the poles of the collective response (\ref{rfcoll}).
One may write
\begin{equation}
\label{zerofreqsec} 
{1\over k} +  \chi (\omega) \approx
    \left({1\over k} +  \chi (0)\right) +\omega \left({\partial
    \chi \over \partial \omega }\right)_{\omega = 0} + \omega ^2 \left(
    {1\over 2}{\partial ^2 \chi \over \partial \omega^2}\right)_{\omega
= 0}
    =0 
\end{equation}
This form invites to define coefficients for friction
$\gamma(0)$ and inertia $M(0)$ by
\begin{equation}
\label{zerofrilim}
\gamma(0)=-i{{\partial \chi (\omega)}\over{\partial \omega}}
    \Bigr\arrowvert_{\omega = 0}={{\partial \chi^{\prime \prime} 
(\omega)}
    \over{\partial \omega}}    \Bigr\arrowvert_{\omega = 0}  
\end{equation}
and
\begin{equation}
\label{zeromass}
M(0)={1\over 2}{{\partial^2 \chi (\omega)}\over{\partial \omega^2}}
    \Bigr\arrowvert_{\omega = 0}={1\over 2}{{\partial^2 \chi^{\prime} 
(\omega)}
    \over{\partial \omega^2}}    \Bigr\arrowvert_{\omega = 0}  
\end{equation}
which in the past have been called the coefficients in "zero frequency"
limit. The effective stiffness $C(0)$ is seen to be identical to the
local stiffness of the quasi-static energy (mind (\ref{appa})), as one 
would
expect to hold true for slow motion, indeed. Incidentally, the
expression for the inertia can be shown to be a generalization of the
one of the cranking model to the case of damped motion \cite{holet}. 
Unfortunately, for strong damping this $M(0)$ becomes very small, and
sometimes even negative. For practical applications this does not
always lead to problems, as usually damping may be so strong that
inertia
must drop out of the macroscopic equations of motion. Nevertheless, this
behaviour of the inertia $M(0)$ is very unpleasant, but fortunately,
one can do better.

Rather than to concentrate just on the denominator of (\ref{rfcoll}), it
is better to take into account the full information contained in this
expression.  As we may recall this was done implicitly when constructing
the transport coefficients $M(\omega_1), \gamma(\omega_1)$ and
$C(\omega_1)$ by approximating (\ref{nonmarkeom}) by (\ref{eomosc}). An
equation like (\ref{eomosc}) may be obtained by expanding $(\chi_{\rm
coll}(\omega))^{-1}$ in (\ref{nonmarkeom}) to second order in $\omega$.
In this way one gets
\begin{equation}
\label{colzerstif}
C \approx
    {1\over k^2 \chi_{\rm coll(\omega)}}\Bigr\arrowvert_{\omega=0}= 
    {{\chi (0)+C(0)}\over{\chi(0)}}C(0) 
\end{equation}
\begin{equation}
\label{colzerfric}
\gamma \approx {1\over k^2}
    {\partial (\chi_{\rm coll(\omega)})^{-1} \over \partial \omega}
    \Bigr\arrowvert_{\omega=0}=
    {{(\chi (0)+C(0))^2}\over{\chi^2(0)}}\gamma(0)    
\end{equation}
and
\begin{equation}
\label{colzermass}
M \approx {1\over 2k^2}{\partial^2 (\chi_{\rm coll(\omega)})^{-1} 
    \over \partial \omega^2} \Bigr\arrowvert_{\omega=0}= {{(\chi 
(0)+C(0))^2}
    \over {\chi^2(0)}} \left(M(0)+{\gamma^2(0)\over \chi (0)}\right) 
\end{equation}
As compared to the zero frequency limit defined above, there are two
modifications: Whereas all three coefficients obtain a factor of
proportionality, only the inertia gets an additional contribution. In
most practical applications this proportionality factor is not very
important, as usually the static response is much larger than the
static stiffness: $\chi(0) \gg \mid C(0)\mid $. It is only for small
excitations and for deformed oscillator shell models, in particular,
that the size of $\mid C(0)\mid $ becomes comparable to that of
$\chi(0)$. However, the additional term in (\ref{colzermass}) ensures 
that the modified inertia does not drop indefinitely anymore with 
increasing
damping. Later on we will demonstrate with the help of
numerical results that (\ref{colzerfric}) and (\ref{colzermass}) 
approximate the selfconsistent friction $\gamma(\omega_1) $ and
inertia $M(\omega_1)$ very well at temperatures $T\ge 2 MeV$. 
To distinguish from the zero frequency limit (\ref{zerofrilim}) and
(\ref{zeromass}) we will associate the approximation (\ref{colzerfric}) 
and
(\ref{colzermass}) to "the zero frequency limit for the collective 
response 
function".

Finally, we should like to mention that relations similar to the ones
given in (\ref{colzerstif})-(\ref{colzermass}) were obtained earlier in
\cite{hosaya}, namely for the model case that the collective response 
function
just consists of one (approximately Lorentzian) peak. Solving the
equations for $C(0),\gamma (0)$ and $M(0)$ given in \cite{hosaya} with
respect to $C, \gamma $ and $M$ one gets back to the equations
(\ref{colzerstif})-(\ref{colzermass}).  

\subsection{ Transformation to sharp densities}

In the previous section we found microscopic expressions of transport
coefficients for large scale motion. For many reasons it is desirable
to compare them with those of "macroscopic models" \cite{swiatcop}, 
like there are the liquid drop model for the inertia and the wall
formula for dissipation---not to mention stiffness, which we
have seen to become identical to the one of the static energy anyway,
as soon as collective motion becomes sufficiently slow. In \cite{hoivya}
many points of principle nature have been clarified about how the
macroscopic limit can be obtained in microscopic theories, concentrating
largely on vibrations around stable configurations. Here we like
to look at another, more practical, albeit very important issue.

By its very nature, these macroscopic models assume the nuclear density
to be constant inside some surface at which the density drops from the
nuclear matter value down to zero within zero range. Commonly this
surface is parameterized with the same set of shape variables which in
our description define the deformation of the mean field, for which in
the present discussion we have chosen the one collective coordinate
$Q$. Contrary to the macroscopic picture, the microscopic calculation
leads to a density distribution with a soft surface, which in addition 
depends essentially on the occupation of states. Take the simple case
of a spherical potential: if a shell with fixed angular momentum $l$ 
is not filled completely the corresponding density distribution will
not be spherically symmetric. On the other hand, it is clear that 
the transport coefficients $\cal T$ will depend sensibly on the
distribution of matter. Thus one needs to employ some specific
transformation to relate them from one case to the other. This is best
done looking at the mean value (or moment) of the operator $\hat F$, as
a representative for the average density.

In the case of our microscopic picture, we have seen the
self-consistency condition (\ref{selfcond}) to imply the relation ${\cal
T}^F=k^2{\cal T}$ between the transport coefficients of the $F$ and
$Q$-mode. All we need to do is to search for a similar condition which
translates the $F$-motion into that of the sharp density distribution,
with the latter being expressed through the corresponding shape
parameter. Such a relation can be found by applying the hypothesis,
which we will substantiate below, that the average $\big\langle F
\big\rangle_{sharp}$, calculated with a density distribution having a 
sharp surface, can be approximated well  by
applying an appropriate Strutinsky smoothing to the shell model
density. The latter is defined as 
\begin{equation}
\label{strmom}
\big\langle F \big\rangle_{sharp} \approx
    \sum_j F_{jj}\tilde n_j    
\end{equation}
where the $\tilde n_j$ are smoothed occupation numbers 
\cite{strut}, \cite{funnyh}.  How they may be used not only to
calculate static expectation values like in (\ref{strmom}), but of
corresponding response functions as well has been studied in
\cite{hofiva} and \cite{hoivya}.  The desired relation between
$\big\langle F \big\rangle_{sharp}$ and $Q$ may now be obtained simply
by applying to (\ref{strmom}) the derivative with respect to $Q$. One
finds
\begin{equation}
\label{chigamma}
{\partial \big\langle \hat F \big\rangle_{sharp}\over \partial Q} 
    \approx {\partial \over \partial Q}\sum_j F_{jj}\tilde n_j=
    \sum_{jk}{\tilde n_j-\tilde n_k\over\epsilon_j-\epsilon_k} 
    \vert F_{jk}\vert^2
    +\sum_j{\partial \tilde n_j\over \partial Q}
    {\partial \epsilon_j\over\partial Q} \equiv -\chi^{\gamma}   
\end{equation}
which may be used to deduce ${\cal T}_{sharp}=(\chi^{\gamma})^2$ ${\cal
T}^F_{sharp}$. Here,  ${\cal T}_{sharp}$ represents the transport
coefficients for the sharp density distribution, but calculated for 
the associated $Q$-mode. Combining this relation with the previous one,
we get  
\begin{equation}
\label{qtransp}
{\cal T}_{sharp}
    =(\chi^{\gamma})^2{\cal T}^F_{sharp} \approx 
    (\chi^{\gamma})^2{\cal T}^F(\omega_1)=(k\chi^{\gamma})^2{\cal 
T}(\omega_1) 
\end{equation}
As the only one further approximation we have assumed that both
density distributions lead to the same averaged value of the field
operator $\hat F$. In (\ref{qtransp}) there appear on the very left
and on the very right the transport coefficients for the $Q$-mode, once
for the sharp surface of the macroscopic models and once for the
collective coordinate specifying the mean field. We may add here that
the quantity $\chi^{\gamma}$ has a physical meaning similar to
that of a static susceptibility, hence the choice of this symbol. The
only difference to the isothermal susceptibility is found in using
the smoothed occupation numbers $\tilde n_j(e_j)$ instead of
$n_j(e_j;T)$.

Let us turn now to "prove" the hypothesis made. This can be done
explicitly for the simple case of the  deformed oscillator potential.
For the more general case we want to appeal to physical intuition, to
the extent of accepting the idea that by its very construction
Strutinsky smoothing commonly does lead to the macroscopic picture.

Suppose we are given a spheroid whose deformation $Q$ is fixed by the
ratio between the semi-axes in $z$-direction and the one perpendicular
to it: $Q=z_0/y_0=\omega_{\perp}/\omega_z$ The sharp surface $S$ is then
given by
\begin{equation}
\label{spheroids}
{x^2+y^2\over y_0^2}+{z^2\over z_0^2}=1    
\end{equation}
The deformation of the sharp density may be classified by the average of 
the
quadrupole operator $\hat Q_2(\vec r)=2z^2-x^2-y^2$ calculated as
\begin{equation}
\label{quamom}
\big\langle Q_2 \big\rangle_{sharp} 
    =\int Q_2(\vec r) \rho_0(\vec r)d\vec r
    ={2A  \over 5}(z_0^2-y_0^2)    
\end{equation}
Here, $\rho_0(\vec r)$ measures the density of homo- geneous nuclear
matter representing $A$ nucleons distributed uniformly within this
surface. In Fig.7 the quadrupole moment (\ref{quamom}) is plotted by the
line with stars as function of the deformation parameter
$Q=\omega_{\perp}/\omega_z$. 

Conversely, the surface (\ref{spheroids}) may be interpreted as an
equipotential surface of the corresponding deformed oscillator potential
\footnote{For this case the quadrupole operator $\hat Q_2$
would be related to our field operator $\hat  F$ taken at spherical
shape by  $F\vert_{Q=1}=-{1\over 3}m \omega_0^2\hat Q_2$.
}
. 
For such a potential one may compute single-particle wave functions
$\varphi_i(\vec r)$ and from them the moment of the microscopic density
as 
\begin{equation}
\label{qudens}
\big\langle Q_2 \big\rangle_{dens}=\int Q_2(\vec r)\sum_j n_j
    \vert \varphi_j (\vec r)\vert^2 d\vec r   
\end{equation}
Obviously this moment $\big\langle Q_2 \big\rangle_{dens}$ depends on
the occupation numbers $n_j$. In  Fig.7 we show curves corresponding to
three choices of $n_j$: Dotted line: the occupation numbers are fixed at
the spherical shape, where the lowest energy states are filled, and kept
constant, independent of $Q$ ("diabatic" situation); dashed curve: at
each crossing of states the particles and holes are redistributed in
such a way that always the lowest states are occupied ("adiabatic"
situation); solid curve: here, the smoothed occupation numbers of the
shell correction method are used.  We see that for the diabatic
occupation numbers the quadrupole moment (\ref{qudens}) differs
substantially from that of the sharp density distribution  $\big\langle
Q_2
\big\rangle_{sharp} $ (line with stars), approximately by a factor of 
two.
Recall that this sharp density distribution reflects the deformation of
the potential.
\begin{figure}[h]
\centerline{{
\epsfysize=8cm
\leavevmode
\epsffile[91 303 482 567]{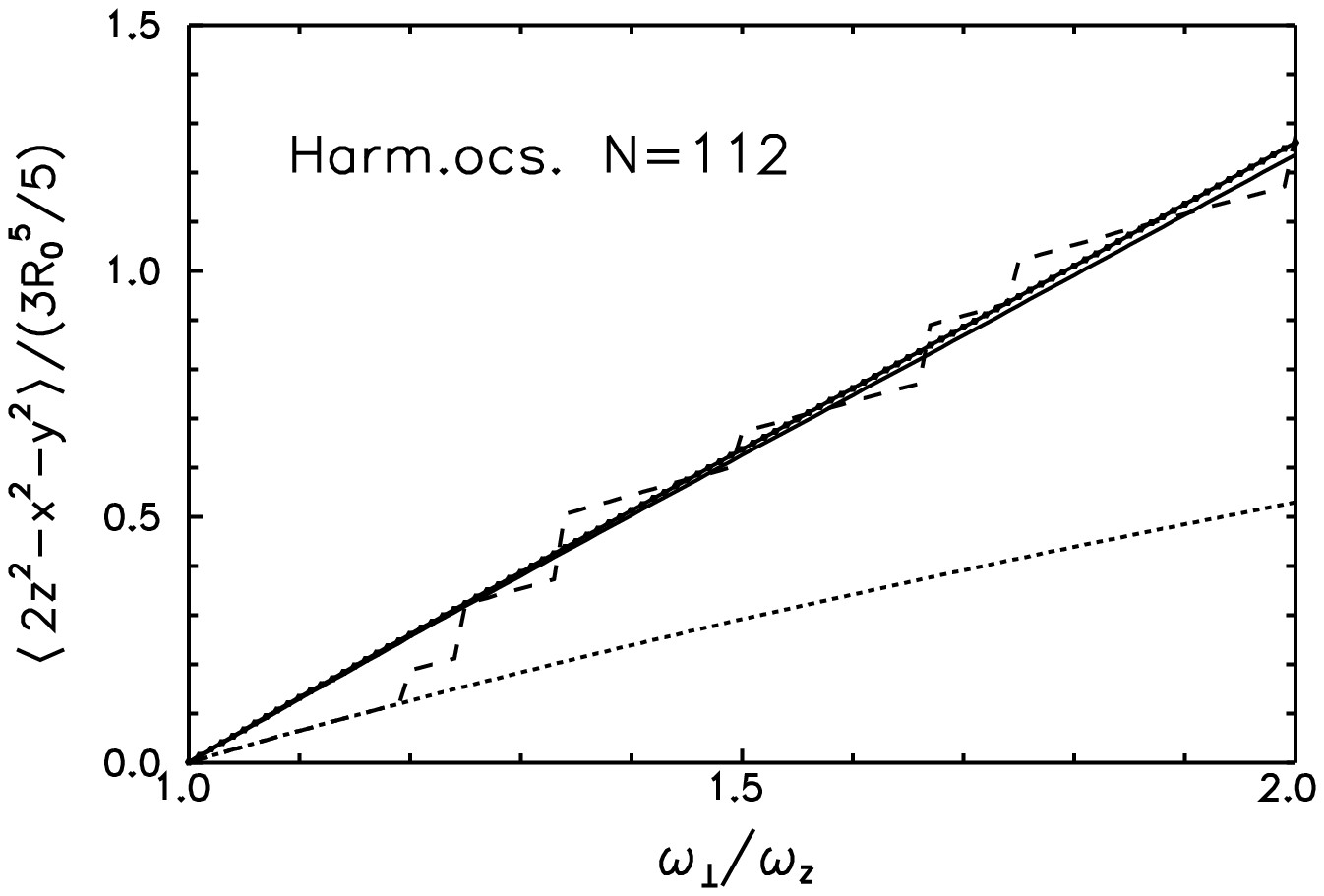}
}}
\caption{A demonstration of the problem of consistency between the
deformation of potential and density, exemplified at the deformed
oscillator; for details see text.
}
\label{Fig.7}
\end{figure}

On the contrary, the density computed with the adiabatic occupation
numbers follows the deformation of the potential {\it on average}.  It
can be said that this feature is one of the basic elements of the
Copenhagen picture of collective motion \cite{bohrmo2} (see also
\cite{sije}).  Indeed, such a redistribution of particles leads to a
consistency between the shapes of the potential and the density. For
static situations, this leads to the well known relation between the
occupation numbers in the various directions and the corresponding
frequencies of the potential. In the treatment of \cite{bohrmo2} this
relation is fulfilled for specific deformations. In Fig.7, the latter
correspond to the points where the broken line crosses that with the
stars. For this broken line it is hardly possible to define precisely
the derivative of $\big\langle Q_2 \big\rangle_{dens}$ with respect to
the deformation parameter. On the other hand, the condition just
mentioned can be fulfilled everywhere using the Strutinsky smoothed
occupation numbers. In this way the derivative is well defined.
Furthermore, the quadrupole moment of the density computed with the
smoothed occupation numbers practically coincides with that of the
sharp surface distribution.

\section{ Microscopic input}

In this section we are going to specify further details of our treatment
of nucleonic dynamics. It has already been mentioned that the mere shell
model is not sufficient. At finite excitations the effects of collisions
cannot be neglected; one even expects them to become the more important
the higher the nucleonic temperature will be.  To treat collisions on
the basis of an explicit form of a two body interaction $\hat
V^{(2)}_{res}(\hat x_i,\hat p_i)$ is hardly possible.  Therefore we
follow another path and parameterize the effect it would have on the
single particle energies. Details of this method can be found in the
publications mentioned before, in particular in \cite{book} (see also
\cite{alkhosi}, \cite{yahosa}, \cite{hoivya}).

\subsection{ Intrinsic response function}

The Fourier transform of the intrinsic response function 
given in (\ref{tdepresp}) can be expressed as the 
sum over single-particle states
\begin{equation}
\label{respfunsum}
\chi (\omega) =\sum_{jk} \chi_{jk}(\omega) 
    \vert F_{jk}\vert^2    
\end{equation}
with
\begin{equation}
\label{respfun}
\chi_{jk}(\omega) =-
   \int_{-\infty}^\infty\;{d\Omega \over 2\pi\hbar }\; n(\Omega )\;
    \Bigl(\varrho_k(\Omega ) {\cal G}_j(\Omega +\omega +i\epsilon)+ 
    \varrho_j(\Omega ) {\cal G}_k(\Omega - \omega -i\epsilon)\Bigr) 
\end{equation}
Here $n(x)$ is the Fermi function determining the occupation of
single-particle levels. The $\varrho_k(\omega)$ represents the
distribution of single-particle strength over more complicated states.
It may be parameterized by
\begin{equation}
\label{dspspecdensgam}
\varrho_k(\omega) = {\Gamma(\omega) \over
    \left(\hbar \omega -e_k -\Sigma^{\prime}(\omega)\right)^2 + 
    \left({\Gamma(\omega)\over 2} \right)^2}
\end{equation}
 in terms of the real and imaginary part of the self-energy
$\Sigma(\omega,T)=\Sigma^{\prime}(\omega,T)-i\Gamma(\omega,T)/2$ 
which are assumed to have the following forms:
\begin{equation}
\label{imselfenomt}
\Gamma(\omega,T)=
    {1\over \Gamma_0}\;{(\hbar \omega - \mu)^2 + \pi^2 T^2 \over 
    1 +{1\over c^2}\left[(\hbar \omega - \mu)^2 + \pi^2 T^2 \right]}
\end{equation}
and
\begin{equation}
\label{reselfenomt}
\Sigma^{\prime}(\omega,T)=
    {-c^2\over 2\Gamma_0\sqrt{c^2+\pi^2 T^2}}\;{(\hbar \omega - \mu) 
\over 
    1 +\left[(\hbar \omega - \mu)^2 + \pi^2 T^2 \right]/c^2}
\end{equation}
Both are connected to each other by a Kramers-Kronig relation. The $\mu$ 
in (\ref{imselfenomt})-(\ref{reselfenomt}) is the chemical potential and 
the cut-off parameter $c$ accounts for the fact that the imaginary part
of the self-energy does not increase indefinitely when the excitations
get away from the Fermi energy. In the present calculation we choose
$\Gamma_0=33 MeV$ and $c=20 MeV$.  The ${\cal G}_k$ appearing in
(\ref{respfun}) is the one-body Green function
\begin{equation}
\label{green}
{\cal G}_k(\omega \pm i\epsilon)={1\over{\hbar \omega
    -\epsilon_k-\Sigma^{\prime} (\omega ,T)  \pm i\Gamma(\omega ,T)/2}} 
\end{equation}
which is related to the spectral density $\varrho_k$ by
\begin{equation}
\label{rhogreen}
\varrho_k(\omega)= i({\cal G}_k(\omega +i\epsilon)
-{\cal G}_k(\omega -i\epsilon))
\end{equation}
Details about the evaluation of the integral in (\ref{respfun}) 
are given in the Appendix.

For future purpose we want to use this form (\ref{respfun}) of the 
response
function and write down a more detailed expression for the friction
coefficient in zero frequency limit:
\begin{equation}
\label{zerofric}
\gamma(0) = -\int{d\hbar \Omega\over 4\pi }{\partial n(\Omega)
    \over{\partial \Omega}}
    \sum _{jk } \big\arrowvert F_{jk }\big\arrowvert ^2  
    \varrho_{k}(\Omega)\varrho_{j}(\Omega)     
\end{equation}
It is obtained (see \cite{hoivya}) by substituting
eqs.(\ref{respfunsum}) 
-
(\ref{respfun}) into (\ref{zerofrilim}). 

\subsection{ The problem of ergodicity}

In section 3.1  it was mentioned
that the condition of ergodicity (\ref{ergodic}) is hard to fulfill in 
the
deformed shell model. This statement refers to the study presented in 
\cite{hoivya}, where it was shown that even collisional damping does not
help, at least in the version as used to date. One important reason was
seen to lie in the fact that our renormalized single particle energies
have the same degeneracies as those of the pure shell model. But these
degeneracies are by far bigger and happen much more often than one
would expect for configurations of the compound nucleus. If one
believes the latter to be important 
--- which should be the case for a fissioning system which has to
overcome a large barrier ---
one needs to take special measures to cure the problem just mentioned. 

One way to make the deficiencies apparent is to look at the heat pole
which shows up either in the correlation function $\psi^{{\prime
\prime}}(\omega)$, or in the relaxation function $\Phi^{\prime\prime}
(\omega)$. Both are related to the dissipative response as
\begin{equation}
\label{correlfun}
\chi^{{\prime \prime}} (\omega) = {1\over \hbar}\tanh
  \left({\hbar\omega\over 2T}\right)\, \psi^{{\prime \prime}} (\omega) 
  \qquad\qquad \Phi^{{\prime \prime}} (\omega) 
  ={\chi^{{\prime \prime}} (\omega)/\omega}
\end{equation}
(see \cite{hoivya} or \cite{book} for more details as well as for
references to the original literature). Both functions have a peak at
$\omega=0$, whose width $\Gamma_T$ was seen to be twice the single
particle width (\ref{imselfenomt}) calculated at the chemical potential,
i.e. $\Gamma_T = 2 \Gamma(\omega=\mu,T)$. On very general grounds, the
height of this peak can be seen to be proportional to the difference of
isothermal and isolated susceptibilities, $\chi^{\rm T} -
\chi(0)$. Numerical calculations in \cite{hoivya} showed this height to 
be
large. Rewriting this difference as  $(\chi^{\rm T} - \chi^{\rm ad}) +
(\chi^{\rm ad} - \chi(0))$ and recalling that in the nuclear case the
difference between the isothermal and adiabatic susceptibility,
$(\chi^{\rm T} - \chi^{\rm ad})$, is small \cite{kidhofiva},
\cite{hoivya}, the origin is identified to come from a large violation
of ergodicity: $\chi^{\rm ad} \neq \chi(0)$.

This discussion indicates what we can do to cure this problem: Cut the
contributions of the heat pole to all functions mentioned previously
down to the magnitude it would have in case the system were ergodic.
This means to reduce the height of this peak at $\omega=0$ by the factor
$(\chi^{\rm T} - \chi^{\rm ad}) / (\chi^{\rm ad} - \chi(0))$.  In
\cite{hoivya} a system was studied were $\chi^{\rm T} - \chi^{\rm ad}$
vanishes identically, such that the reduction of the heat pole amounted
to neglect contributions from all states having the same energy. Such a
correction can easily be done. In expressions like (\ref{respfunsum})
one simply has to restrict the summations in a proper way. Here we like
to adopt a similar procedure, in the sense of neglecting the influence
of a finite difference $\chi^{\rm T} - \chi^{\rm ad}$. At the
temperatures considered it will be very small, indeed. On the other hand
we go one step further and neglect also contributions from neighboring
states whose energy is finite but smaller than the collisional width the
particles, which as we just saw also reflects the width of the heat
pole.

The consequences of such a manipulation are shown in Fig.8 for the
intrinsic response (left part), the corresponding relaxation function
$\Phi^{{\prime \prime}} (\omega)$ (upper right part) and the collective
strength distribution $\chi^{\prime\prime}_{\rm coll}(\omega)$ (lower
right part).  All of them have been computed for $T=1\,MeV$.  The fully
drawn lines correspond to calculations where all matrix elements are
taken into account. For the dashed curves matrix elements $F_{jk}$
between states of energy difference $|(e_k-e_j)|\leq \Gamma (\mu, T)$
have been discarded. Their contribution to the relaxation function is
exhibited in the upper right part by the dotted line. This plot
demonstrates nicely that the heat pole can be associated to a Lorentzian
of width $\Gamma_T$ around $\omega=0$, and thus corresponds to a pole on
the  imaginary axes \cite{hoivya}, \cite{book}.
\begin{figure}[h]
\vspace*{0.5cm}
\centerline{{
\epsfysize=10cm
\leavevmode
\epsffile[71 266 490 530]{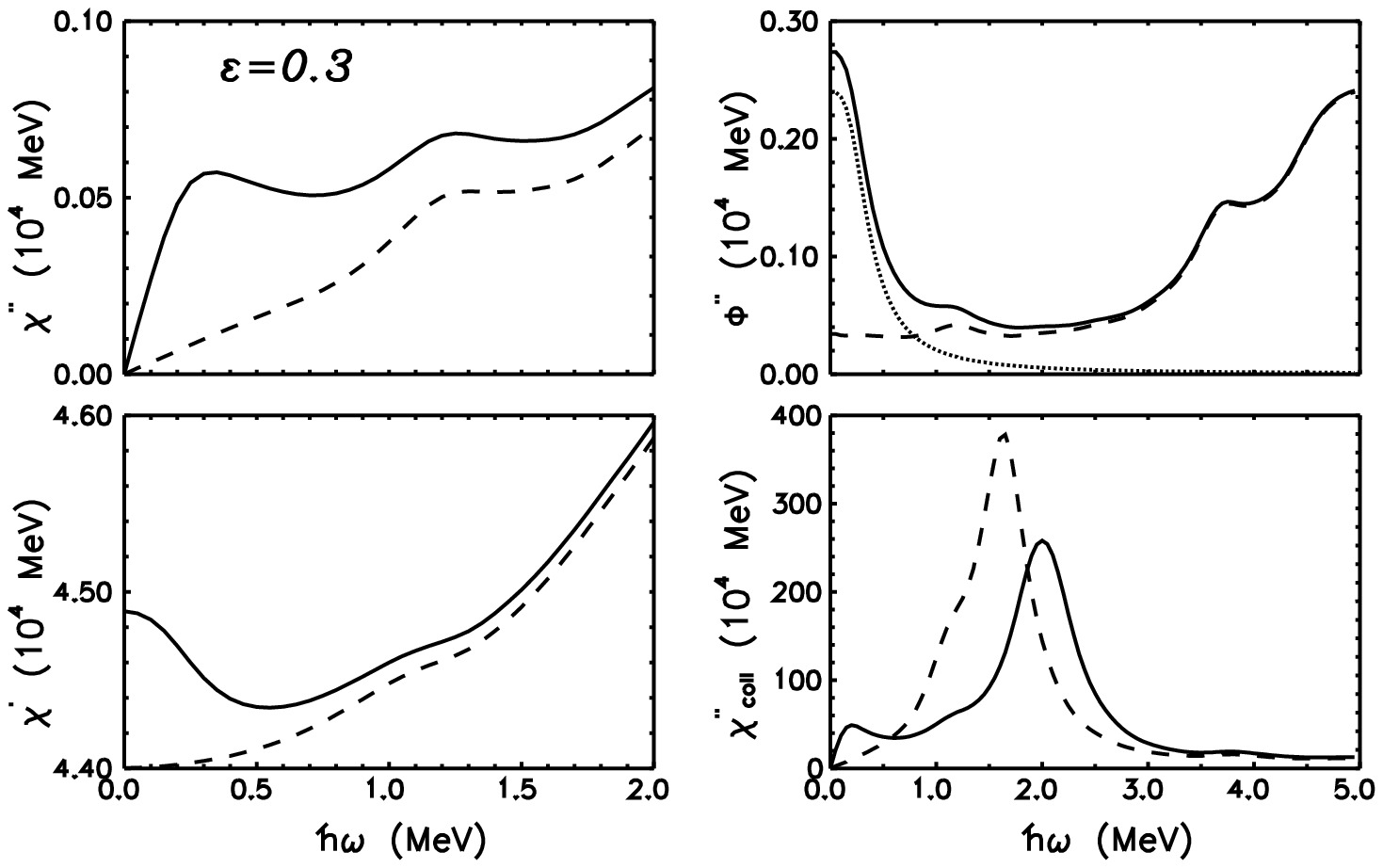}
}}
\vspace*{-0.5cm}
\caption{The heat pole contribution to the relaxation and
response functions, calculated at $T=1\,MeV$. 
}
\label{Fig.8}
\end{figure}
It is interesting to note that for the present calculation $99 \%$ of
this Lorentzian is made out only by two {\it pseudo-crossings} of
single-particle states which takes place close to the Fermi energy. By
pseudo-crossing we mean a situation where as function of $Q$ two levels
come close but never cross; one such event is encircled in Fig.3. The
big contribution to friction in the zero frequency limit of
(\ref{zerofrilim}) (or (\ref{colzerfric})) which results from this heat
pole can be estimated looking at the upper right part of Fig.8: The
slope of the fully drawn line is much bigger than the one of the dashed
line. On the other hand, contributions to $\gamma(0)$ from {\it real
crossings} of levels is very small. This is due to the fact that the
matrix elements $|F_{jk}|$ vanish exactly at the crossing points and are
very small in the vicinity of such a (real) crossing.
  
The influence of the heat pole (in the intrinsic system) on the
collective strength distribution can be inferred from the lower right
part of Fig.8. The fully drawn line shows a small peak at very small
frequencies, say $\hbar \omega \approx 0.2-0.3 MeV$. Compared to the
much bigger strength found in the peak at about $1.5\;-\;2 \;MeV$ one is
inclined to just "forget" the small peak when one wants to define the
transport coefficients. Indeed, it somehow looks very natural to
associate the bigger peak to the genuine low frequency mode.  This may
be understood as another argument to leave out the contribution of the
heat pole to the transport coefficients, besides the one involving
ergodicity for intrinsic motion. All computations to be reported below
were done along this line, i.e. contributions to response function from
the states with the energy difference $|(e_k-e_j)|\leq \Gamma (\mu, T)$
were not taken into account.

\subsection{ The influence of collisional damping on nucleonic 
relaxation}

The whole formulation of our theory is based on the assumption that the
nucleonic degrees of freedom stay close to a thermal equilibrium. The
latter is not fixed, however; rather it continuously gets disturbed by
collective motion itself, forgetting for the moment a possible
evaporation of light particles or gammas. It should thus be of interest
to have some estimate of an appropriate relaxation time.  The best
candidate for this is offered by the "generator" of collective motion,
namely the $\hat F({\hat x}_i, {\hat p}_i, Q_0)$, which defines the
coupling of the collective variable to the nucleons. It is predominately
this quantity which "decides" which kind of modes of the intrinsic
degrees of freedom get excited. We may recall from the discussion in
sect.3 the close relation of this $\hat F$ to the nucleonic response
function appearing in our theory: The $\chi(\omega)$ parameterizes that
average "excitation" $\delta \big\langle \hat F \big\rangle_{\omega}$
which comes about through a change of the  collective variable $\delta
Q$. If we are just interested to estimate the time $\tau$ after which
the $\delta
\big\langle \hat F \big\rangle_{t}$ has decayed to its static value, we
may study the time dependent function $\tilde{\chi}(t)$ given in
(\ref{tdepresp}). In literal sense it represents the $\delta \big\langle
\hat F \big\rangle_{t}$ if excited by a sharp pulse like $\delta Q(t) 
\propto
\delta(t)$. Notice please, that for $t>0$ the $\tilde{\chi}(t)$ is
proportional to the derivative of the Fourier transformed relaxation
function, namely $\tilde{\chi}(t)\propto d\,\tilde\Phi^{{\prime
\prime}}(t)\,/\, dt$ (see (\ref{tdepresp}) and (\ref{correlfun})). Thus
the information contained in $\tilde{\chi}(t)$ is equaivalent to that of
$\Phi^{{\prime \prime}}(t)$ up to an additive constant. The latter
measures the long time limit of $\Phi^{{\prime \prime}}(t)$; it is
related to the strength of the heat pole. At the moment we are
interested only in the behaviour for finite times.

In Fig.9. results of two different computations of $\tilde{\chi}(t)$ are
shown. One case just refers to the deformed shell model, for the other
one collisional damping is taken into account. It is clearly seen that
only by way of such collisions we may speak of genuine relaxation.  It
is also observed that the latter does not depend much on the shape.
However, the relaxation time $\tau$ decreases considerably with
increasing temperature. The latter effect is expected of course from the
very form by which the temperature appears in the single particle widths
(\ref{imselfenomt}). From the fully drawn lines of Fig.9 one may deduce
for $\tau$ values like $0.5,\,0.3,\,0.1\;\hbar/MeV$ for
$T=1,\,2,\,5\;MeV$. 
(This may for instance be done by approximating the envelope of
$\tilde{\chi}(t)$ by an expontial such that $\tau$ may be defined
through the "width" at half maximum.) 
Later on in sect.5.3 we are going to compare them
with typical time scales of collective motion, but we may say already
here that this microscopic time $\tau$ is at least one order of
magnitude smaller than the one which measures motion in the collective
variable $Q$.

Finally, we should like to mention that the results found from the
present computation are in accord with those reported in
\cite{jenlefho}.

\section{Numerical results for collective transport coefficients}
\begin{figure}[t]
\centerline{{
\epsfysize=10cm
\leavevmode
\epsffile[48 224 536 595]{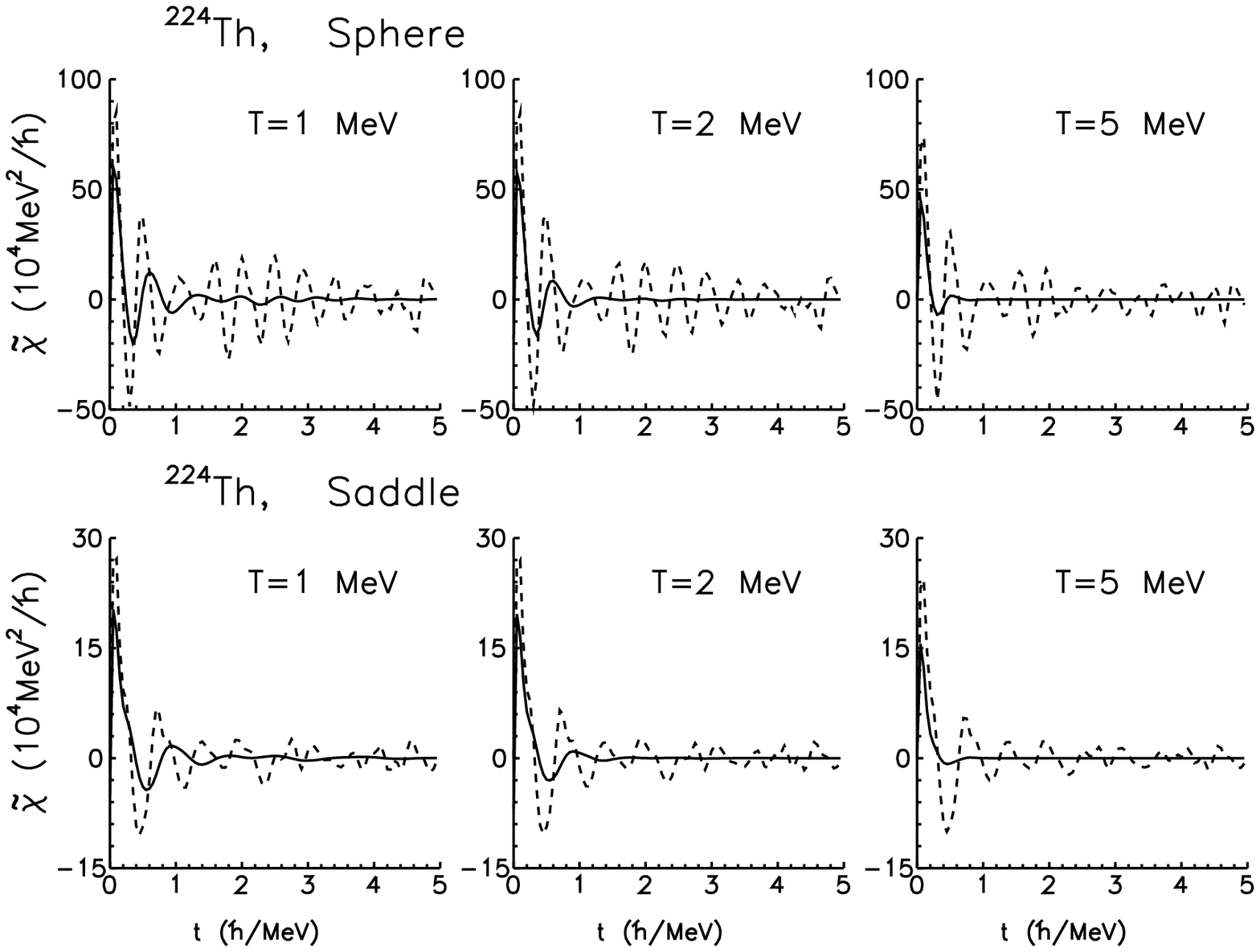}
}}
\caption{The time dependent function for the nucleonic response,
calculated within the pure shell model (dashed curve) and for
collisional damping (fully drawn line).
}
\label{Fig.9}
\end{figure}

In this chapter we will discuss the numerical results for transport
coefficients ${\cal T}$ computed along the fission path of $^{224}Th$.
As it was already mentioned, for temperatures above $1 MeV$ the fission
path is parameterized in terms of Cassini ovaloids by only one
deformation parameter $\epsilon$. Since we prefer to use the parameter
$r_{12}$ instead of $\epsilon$ we have to relate the transport
coefficients accordingly. This can be done by exploiting the following
relation, obeying that the $\alpha_n$'s are fixed,
\begin{equation}
\label{massrr}
{\cal T}_{r_{12}r_{12}}=
    {\cal T}_{\epsilon \epsilon}\left({dr_{12}
    \over d\epsilon}\right)^{-2}  
\end{equation} 
which simply follows from general properties of coordinate
transformations. The derivative $\partial r_{12} /\partial \epsilon$ is
obtained by differentiating (\ref{rr12}), the result is shown by dashed
line in Fig.2. Recall please, that the deformation dependence of the
transport coefficients is defined essentially by the choice of the
collective variables.  For example,  ${\cal T}_{r_{12}r_{12}}$
decreases with $r_{12}$ but  ${\cal T}_{\epsilon \epsilon}$ is increases
as a function of $\epsilon$. Both in the figures as well as in the text
below we will omit the indices $r_{12} r_{12}$, keeping in mind  
that the transport coefficients are defined with respect to  $r_{12}$
(even if sometimes they will be shown as function of $\epsilon$).

\subsection{ Accuracy of zero frequency limit for the
collective response function}

The friction coefficient $\gamma (\omega_1)$ and mass parameter
$M(\omega_1)$ defined according to (\ref{lorfit}) are shown in  Fig.10
by solid lines as function of the deformation parameter $r_{12} $ for
temperatures between $1$ and $3\, MeV$. They are compared with
calculations for which the approximation
(\ref{colzerfric})-(\ref{colzermass}) is used for friction and inertia,
respectively. The latter results are marked by dashed curves.  As can be
seen, this approximation is quite accurate for $T\geq 2 MeV$. This
implies that for such temperatures one may avoid the time consuming
computation of the frequency dependence of collective response
function. One may compute friction and inertia directly from
(\ref{colzerfric}) and (\ref{colzermass}). As for the big fluctuations
seen at $T=1\;MeV$ we expect them to become much smaller as soon as
pairing correlations will be taken into account, which shall be subject
of future studies.
\begin{figure}[h]
\centerline{{
\epsfysize=10cm
\leavevmode
\epsffile[57 278 558 595]{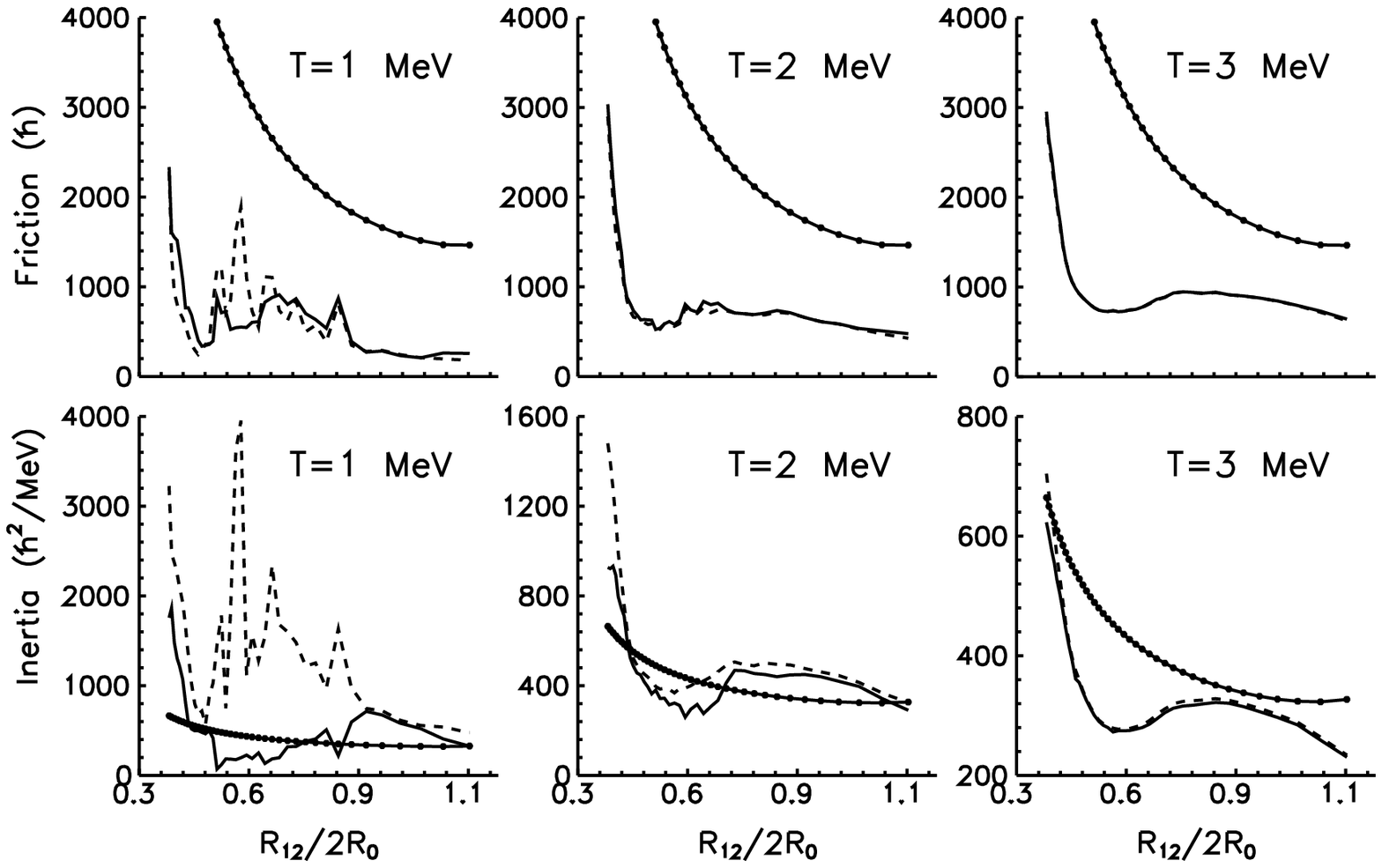}
}}
\caption{The friction coefficient and the mass parameter 
at finite frequency (solid lines) and for the approximation 
(\protect\ref{colzerfric})-(\protect\ref{colzermass}) (dashed lines).
}
\label{Fig.10}
\end{figure}

For comparison we also show in Fig.10
(by the lines with stars) wall friction $\gamma_{\rm
w.f.}$ and the inertia of irrotational flow. 
According to \cite{bbnrrss} wall friction $\gamma_{\rm w.f.}$ is
proportional to the squared normal velocity $u_n^2(s)$ of the surface,
integrated over the nuclear surface. Following \cite{bbnrrss} this may
be deduced from the loss of (collective) energy which is given by
\begin{equation}
\label{wallfric}
\dot E ={3 \over 4}\rho v_F \oint u_n^2(s)ds = \gamma_{\rm w.f.} \dot
r_{12}^2 (t)
\end{equation}
where $\rho$ and $v_F$ are the nucleons' density and Fermi velocity. For
axially symmetric shapes the surface velocity $u_n(s)$ can be expressed
in terms of the profile function $\rho (z, \epsilon)$ from eq.
(\ref{ovals}) as
\begin{equation}
\label{normve}
u_n(z)=\bar u_n(z)\dot r_{12}(t),\,\,\,\,\,
\bar u_n(z)={1\over \Lambda (z)}{\partial \rho (z, \epsilon)\over
\partial \epsilon}\left({\partial r_{12} \over 
\partial \epsilon}\right)^{-1},\,\,\,
    \,\,\,\,\,\Lambda (z)=\sqrt{1+\left({\partial \rho (z, \epsilon)
    \over\partial z}\right)^2}    
\end{equation}
The mass parameter of an incompressible irrotational fluid has been
computed as suggested in  \cite{ivkoma}. It can be written as
\begin{equation}
\label{mirr}
M_{irr}=m \oint \xi (s) \bar u_n(s) ds
\end{equation}
with the potential $\xi (\vec r)$ for the velocity field expressed 
by the potential of some "surface charge" distribution
\begin{equation}
\label{velfield}
\xi (\vec r)={1\over 2\pi} \oint {\nu (s^{\prime}) \over
|\vec r-\vec r(s^{\prime})|}ds^{\prime}
\end{equation}
The substitution of (\ref{velfield}) into the Neumann equation
\begin{equation}
\label{neumann}
\Delta \xi =0,\,\,\,\,(\vec n \nabla \xi)_S=u_n(s)  
\end{equation}
leads to some integral equation for the density of the "surface charge"
$\nu (s)$ which was solved iteratively starting with
\begin{equation}
\label{nuzero}
\nu^0(s)=-\bar u_n(s)
\end{equation}
as zeroth approximation to $\nu (s)$, for details see \cite{ivkoma}.  We
have checked that for the particular case of the shape family
(\ref{ovals}) the Werner-Wheeler method \cite{werwhe} turns out to be a
very accurate approximation to the mass parameter (\ref{mirr}). Both
results coincide within the thickness of the lines in the Figure.

\subsection{ Temperature and deformation dependence of friction and
inertia}

In Figs.11 and 12 the friction coefficient and the mass parameter are
presented as function of the deformation parameter $\epsilon$ for five
different temperatures $T= 1-5\, MeV$. They have been calculated from
the oscillator fit (\ref{lorfit}), but transformed to sharp densities
according to (\ref{qtransp}).  The parameter $\epsilon$  is chosen for
reasons to be given below.  In both figures results of the commonly
adopted macroscopic models, namely $\gamma_{w.f.}$ and $M_{irr}$ are
shown by the lines with stars.  Several observations can be made.
\\ $\bullet$ To some extent the $\gamma_{w.f.}$ and $M_{irr}$ can be
said 
to be reached at the higher temperatures. The very fact that this
statement
is more true for the inertia but less so for friction can be understood
as follows. As shown in \cite{hoivya}, $\gamma_{w.f.}$ may be considered
the macroscopic limit of our model only if such subtleties as
collisional damping are left out. Conversely, the high temperature limit
of the inertia is related to the value of the energy weighted sum, and
the latter is known to be associated to $M_{irr}$. This is true at least
when one treats vibrations within simple models (see \cite{bohrmo2},
\cite{sije} for the situations of $T=0$ and \cite{hoyaje} at $T\neq 0$).
\\ $\bullet$ As has been demonstrated in \cite{yaivho}, in the general 
case the
evaluation of the sum rule value becomes somewhat delicate for
collisional damping. The nice feature of reaching the  $M_{irr}$
automatically can be considered a proof of the usefulness of the
transformation (\ref{qtransp}).
\\ $\bullet$ Like in the calculations of \cite{yaivho}, on average 
friction is
seen to increase with $T$ whereas the inertia decreases.
\begin{figure}[t]
\centerline{{
\epsfysize=8cm
\leavevmode
\epsffile[88 286 485 544]{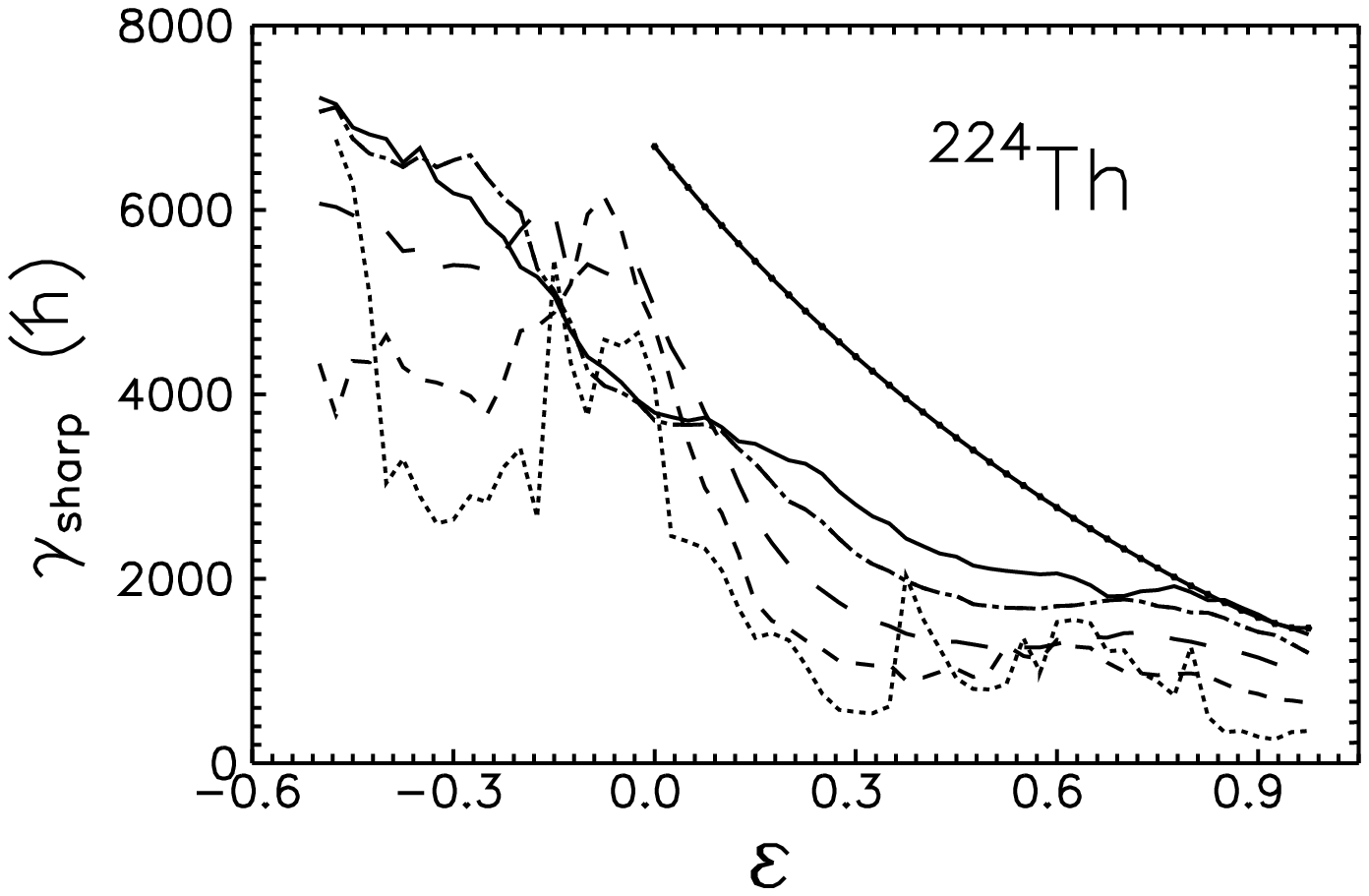}
}}
\caption{The deformation dependence of friction;
the dotted, short-dashed, long-dashed, dotted-dashed and 
solid lines correspond to
temperatures $T=1,2,3,4$ and $5 MeV$.
}
\label{Fig.11}
\end{figure}
\\ $\bullet$ Again similarly to \cite{yaivho},
both transport coefficients decrease with deformation on average.
\\ $\bullet$ The coordinate dependence shows fluctuations around the
average trend, which are more pronounced at smaller temperatures. They
appear to be even larger than those reported in \cite{yaivho}, but they
are similar to the results of \cite{ivapom}.  As for the friction
coefficient, the dependence on deformation is particularly strong around
the spherical shape. It reaches a kind of local  maximum there, or
perhaps at a slightly oblate shape. These features can be exhibited more
clearly for parameterization in terms of $\epsilon$, for which reason we
have made this choice here.

Many of these features point to the importance of shell effects, in
particular the peak for friction around the spherical configuration,
which is clearly visible for $T\approx 1-2 MeV $ but which disappears at
$T\approx 4-5 MeV$. We like to elaborate on this statement by trying to
split the friction coefficient up into a smooth and a fluctuating part.
For this study we take the zero frequency limit.

Let us suppose for a moment that the matrix elements $|F_{jk}|^2$
considered as function of the single-particle energies have some smooth
average component ${\cal F}^2(e,e^{\prime})$ and the oscillating
component can be neglected. In this case one may rewrite
eq.(\ref{zerofric}) in the form
\begin{equation}
\label{zerofrappr}
\gamma(0) \approx -\hbar \pi {\cal F}^2(\mu , \mu)
    \int d\Omega {\partial n(\Omega) \over{\partial \Omega}} g^2(\hbar
\Omega)
\end{equation}
The $\varrho_k(\Omega)$ in (\ref{zerofric}) are peaked functions with
their maximum at $\hbar \Omega =\epsilon_k $ so that the sum of
$\varrho_k$ over $k$ may be interpreted as the density $g(\hbar \Omega)$ 
of single-particle states. As usual the latter can be split into the
smooth
and oscillating components $g(e)=\tilde g(e) +\delta g(e)$ with $\tilde
g(e)=\big\langle g(e) \big\rangle$ and $\big\langle \delta g(e)
\big\rangle =0$ where the brackets stand for an averaging over the
single-particle spectrum.  Inserting this decomposition into
(\ref{zerofrappr}), and noting that an integration performed with the
bell-like function $\partial n(\Omega)/\partial
\Omega$ can be understood like the average introduced above, we will get 
\begin{equation}
\label{zerofrapprg}
\gamma(0) \approx \hbar \pi\left[\bar g^2(\mu)
    +\big\langle \delta g^2(e)\big\rangle_{e=\mu}\right] {\cal F}^2(\mu,
\mu)
\end{equation} 
\begin{figure}[t]
\centerline{{
\epsfysize=8cm
\leavevmode
\epsffile[85 286 485 547]{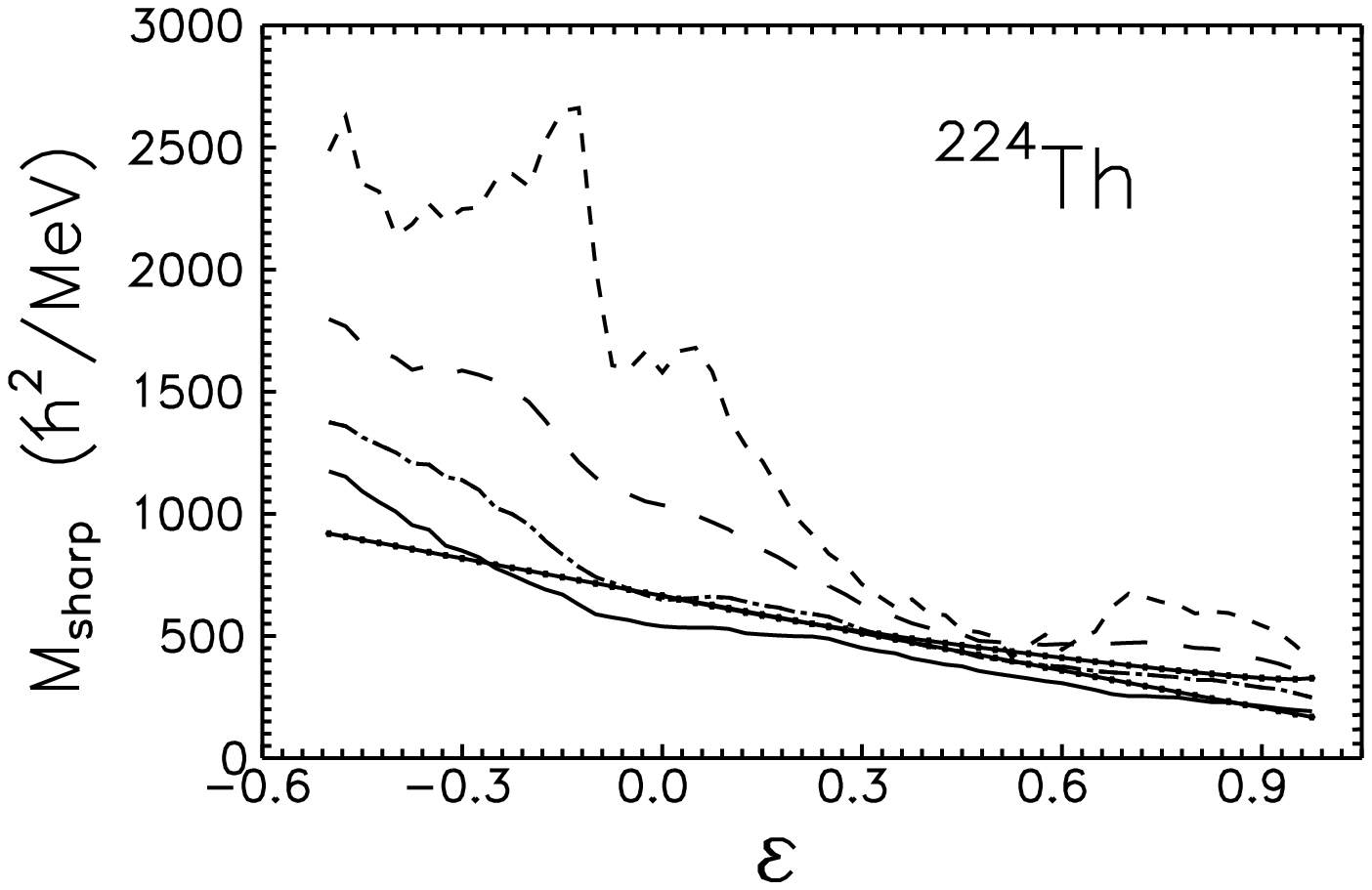}
}}
\caption{The deformation dependence of the mass parameter for
temperatures  $T=1-5 MeV$. Different temperatures are marked by the same 
lines as in Fig.11.}
\label{Fig.12}
\end{figure}
Assuming the oscillating component of the density to be periodic in the
energy with some period $\hbar \Omega_0$ and amplitude $\delta g_0$, viz
$\delta g(e)=\delta g_0\sin{2\pi e/\hbar \Omega_0}$, it is easy to
convince oneselves that $\big\langle \delta g^2(e)\big\rangle
=(1/2)\delta g_0^2$.  The quantity  $\delta g_0$ is determined by the
magnitude of the shell correction. It is a smooth function of particle
number ( see
\cite{ivastr}) but still depends on deformation.  It is maximal at that
deformation where the shell structure is more pronounced. For the
Woods-Saxon potential this happens to be so at the spherical shape. So
it is due to the shell structure that the friction coefficient gains
additional contributions around the sphere. For the
case of the Woods-Saxon potential this specific feature seems to be
responsible for the dip one sees in Fig.10 both for friction as well as
for the inertia around $r_{12}\approx 0.5$, even at the somewhat larger
temperatures of $ 3\; MeV$. It may be said, however, that such
a behavior is not seen in computations performed with the model of
\cite{yaivho}. There the change of the friction coefficient with
$r_{12}$, for instance, resembles more the smooth one given by the wall
formula. 

\subsection{ Collective relaxation times}

It has become customary to parameterize friction in terms of the ratio
$\beta= \gamma / M$, although most of the interest has concentrated on
friction alone; often the inertia was simply taken to be the reduced
mass of the fissioning system. As we have seen above, for a microscopic
theory both quantities will vary not only with the collective variable
but with temperature as well.  It may be expected, of course, that for
the ratio the dependence on shape is much weaker than that of the
individual quantities. This follows simply from the observation that
any common, purely geometrical factor will drop out. In Fig.13 we
present the results for $\beta$ obtained from those for inertia and
friction discussed before. Indeed, this quantity is essentially
constant over the whole deformation region, for all computations but
$T=1\;MeV$, a case for which the fluctuations seen in our results are
too big because of our neglecting pairing.  However, there is a marked
dependence on excitation: $\beta$ increases strongly with $T$. This is
in clear distinction to the result one gets from applying the wall
formula for friction and that of irrotational flow for the inertia.
Interestingly enough, these macroscopic estimates lead to some
$Q$-dependence, which in a sense is even bigger than suggested by the
trend of our results. It can be said that the latter are very close to
those obtained in the computations with a two-center shell model
potential \cite{yaivho}.
\begin{figure}[h]
\centerline{{
\epsfysize=8cm
\leavevmode
\epsffile[111 286 533 544]{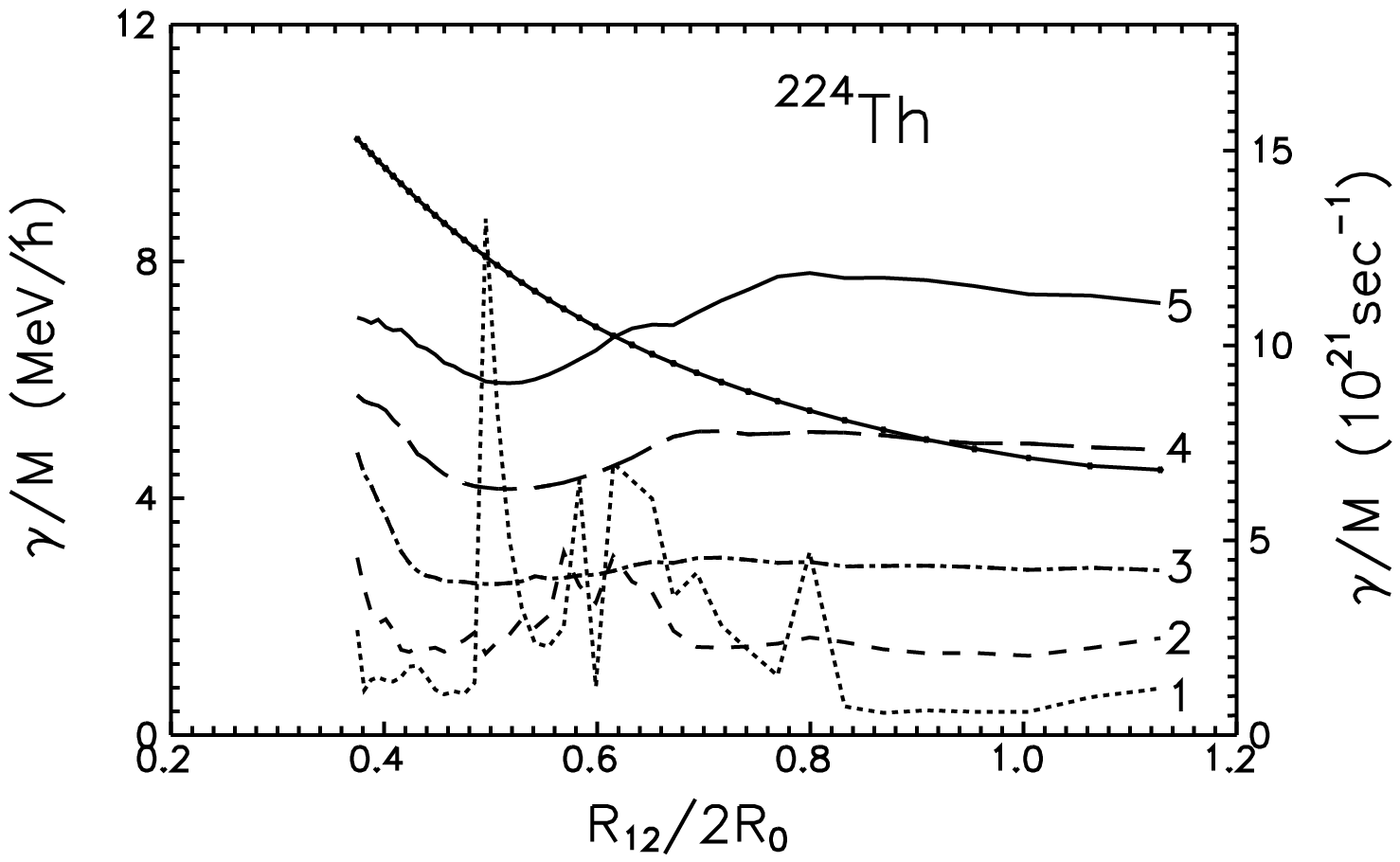}
}}
\caption{The inverse relaxation time $\beta =\gamma 
(\omega_1)/M(\omega_1)$
as function of deformation and temperature (indicated in the Figure).}
\label{Fig.13}
\end{figure}

Physically, the inverse of $\beta$ can be interpreted as the relaxation
time ($\tau_{kin}=M/\gamma$) to the Maxwell distribution for collective
motion. In full glory this feature can only be understood looking at the
dynamics in collective phase space (see e.g.\cite{book}). However, one
may grasp its content recalling the local equation for average motion of
the damped oscillator: $M\ddot q(t) +
\gamma \dot q(t) + Cq(t)=0$. From this equation it also becomes apparent
that yet another relaxation time can be defined, namely $\tau_{coll}=
\gamma/|C|$.  For overdamped motion the latter is the only relevant one.
Actually, such a situation is given for temperatures above $1-2\;MeV$.

Let us infer this feature comparing the macroscopic relaxation times
with the microscopic $\tau$ found in sect.4.3. There we got values in
the range between $\tau=0.3$ and $0.1\;\hbar/MeV$ for temperatures
between $2$ and $5\;MeV$, discarding the case of $T=1\;MeV$ for the
moment.  Taking for the $\beta$ of Fig.13 typical values like $2$ and
$8\;MeV/\hbar$ for $T=2$ and $5\;MeV$, respectively, we recognize the
$\tau_{kin}$ to be comparable or only slightly larger than $\tau$. This
means that in this range of temperatures the friction force is so strong
that it leads to an instant damping or disappearance of the kinetic
energy. This is not in contradiction to the basic assumption behind the
quasi-static picture. To justify the latter all one needs to have is the
motion in $Q$ to be slow compared to that for the intrinsic degrees of
freedom.  For this question it is the $\tau_{coll}$ which becomes
relevant. Its value can be estimated with the help of Figs.6 and 10. Let
us concentrate on deformations for which the local stiffness does not
get zero. Since for $T\ge 2\;MeV$ shell effects are not important
anymore we may estimate $C$ as the one of the liquid drop energy. The
right part of Fig.6.  tells us the $|C_{LDM}|$ to be of the order of
$150\; MeV$ for $r_{12}$ larger than about $0.6$. In this range the
value of $\gamma$ is about $800\;\hbar$, as seen from Fig.10 (for
$T=2\;MeV$; it is even bigger for larger temperatures). For
$\tau_{coll}$ this implies values of the order of $5\;\hbar/MeV$, which
are larger than $\tau$ by more than one order of magnitude.
\section{ Summary and conclusions}

In this paper we have applied the single particle model of \cite{pash71}
to describe large scale motion at finite excitations. To this end this
model had to be modified to include effects of collisional damping and
it had to be adapted to the formulation of collective motion in the
spirit of the locally harmonic approximation (for a review see
\cite{book}). Numerical computations have been performed for the 
transport coefficients of average motion along a fission path. The
latter was identified by the valley in the potential landscape obtained
for the liquid drop model. It was argued that for the range of
temperatures considered in the present study this path may be expected
to represent fairly well the actual situation. First of all, because of
the evidence one has from static considerations that Cassini ovaloids
describe well the shapes of the fissioning nucleus. Secondly, for the
large damping one expects to be given, the system will be creeping down
the collective potential and thus will stay close to the line of
steepest descent.

For the transport coefficients values were found which are in accord
with previous studies, in particular with the ones of \cite{yaivho}.
This is especially so for their dependence with temperature, and to
lesser extent for their variation with the nuclear shape. For instance,
it turns out that typical effects of single particle motion become more
apparent here than they did in \cite{yaivho}, like there are
fluctuations of both friction and inertia with the collective variable.
To large amount they disappear when building ratios like for the $\beta
=\gamma/M$ or the $\eta= \gamma/(2\sqrt{M|C|}$, two quantities which are
commonly used to parameterize collective dynamics.

A new development has been achieved with respect to comparisons with
macroscopic models, like that of irrotational flow for the inertia and
that of wall friction. As one knows, both are calculated for sharp
density distributions which may differ considerably from those which
correspond to the nucleons' densities in the shell model. A
transformation was suggested which allows one to connect both density
distributions and thus enables one to connect the transport coefficients
accordingly.  This transformation was found by applying a Strutinsky
smoothing procedure to evaluate the average static value of the relevant
one-body operator $\hat F({\hat x}_i, {\hat p}_i, Q_0)$. It is this
operator which by a self consistency argument is related to the
deformation of the mean field.

Finally, we briefly like to turn to comparisons with experimental
findings. First of all, we may mention that the values for $\beta$
shown above concur with the range suggested by fission 
experiments \cite{higoroobn}. Two other relevant parameters are the 
$\varpi= \sqrt{|C|/M}$ and the $\eta$ mentioned above, which for
instance appear in Kramers' famous formula for the decay rate of a
one-dimensional fission model: 
${\rm R_K} = \left(\sqrt{1 + \eta_s} - \eta_s\right) \;
(\varpi_m / 2\pi)\, \exp(-B/T) $. Here, $B$ measures the barrier
height, $\varpi_m$ determines the vibrational frequency in the
potential minimum and $\eta_s$ is to be evaluated at the saddle.  
\begin{figure}[h]
\centerline{{
\epsfysize=7cm
\leavevmode
\epsffile[65 275 502 547]{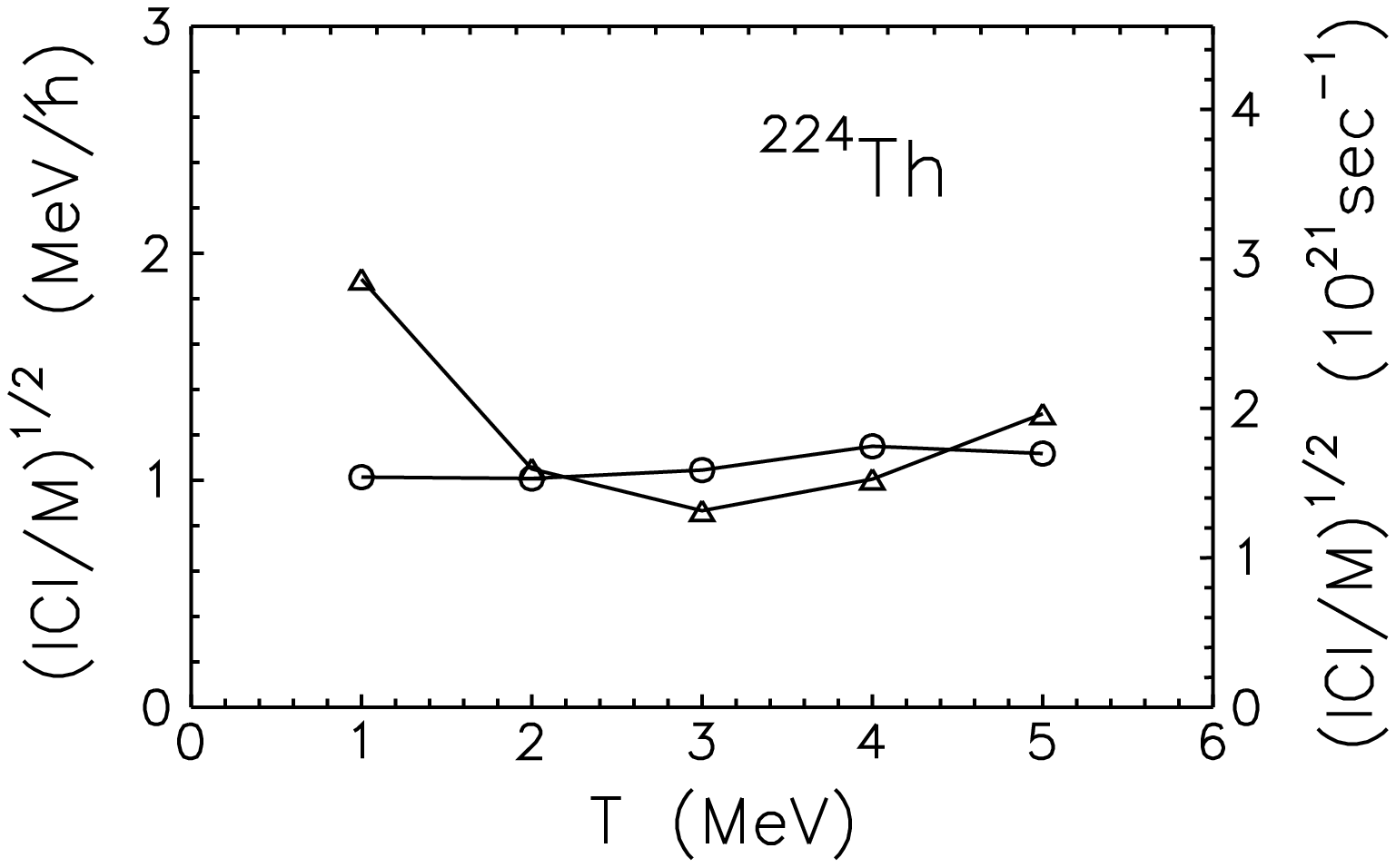}
}}
\caption{The vibrational frequency $\varpi= \protect\sqrt{|C|/M}$ 
as function of
temperature, calculated at the potential minimum (marked by circles) 
and at the saddle point (marked by triangles).}
\label{Fig.14}
\end{figure}
The latter two quantities are shown in Figs.14 and 15, respectively, for
both points. They have been calculated by averaging the
\begin{figure}[h]
\centerline{{
\epsfysize=7cm
\leavevmode
\epsffile[71 278 454 547]{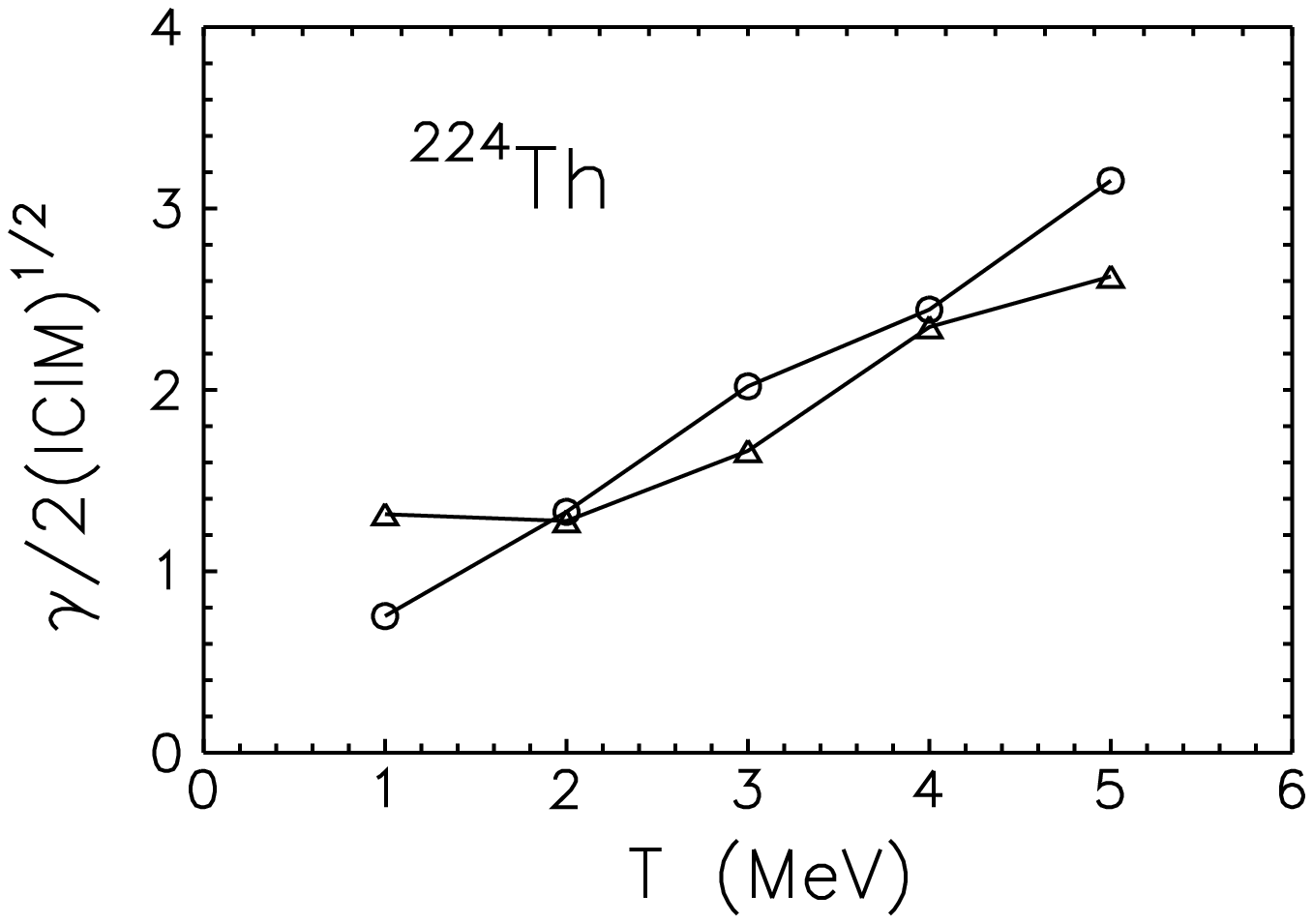}
}}
\caption{The dimensionless parameter 
$\eta =\gamma/(2\protect\sqrt{|C|M})$ as function
of temperature, calculated at the potential minimum (marked by circles) 
and at the saddle point (marked by triangles).}
\label{Fig.15}
\end{figure}
$Q-$dependent transport coefficients in the neighborhood of the
minimum and the barrier, 
where both are those of the $T$-dependent potential energy. 
These results agree very well with
those already shown in \cite{yaivho}. It is remarkable that $\varpi$
does not change much with $T$, neither at the minimum nor at the
barrier. Conversely, like the $\beta$ shown before, the $\eta$ too
definitely increases with temperature. This behavior is in at least
qualitative agreement with the one found in \cite{pauthoe}
and the numbers for $\eta$ lie in the range of values deduced only
recently in \cite{lestone} from experimental evidence. It is true, of
course, that 
for more quantitative analyses one would have to consider both the
effects of
pairing as well as of angular momentum, not to mention the fact that
there is some freedom \cite{book} in adjusting the two parameters which
define collisional damping. However, there is probably little doubt
about the temperature dependence of our coefficients, say above
$T=1.5-2\;MeV$. This marked change with $T$ is in clear distinction to
the macroscopic models mentioned. In the light of this feature
agreements between experimental findings and theoretical descriptions
appear somewhat questionable if they are only based on these
macroscopic models \cite{wada}, \cite{pobaridi}. Likewise the somewhat
peculiar behavior of $\beta$ with $r_{12}$ suggested in
\cite{gonfroe} is not confirmed by our results (see also \cite{yaivho}). 
It is perhaps fair to say that some of the difficulties one still 
encounters at present when comparing theory with experiment is due to
the high complexity of the problem itself as well that of the analysis
of the data.  To underline this statement, we just like to take up a
point raised in
\cite{thoenn}, namely that common statistical codes evaluate the fission
decay rate not by the transition state result, to which Kramers'
formula reduces to for $\eta_s \to 0$, but by that of the "statistical 
model"
where $\varpi_m$ is replaced by $T$. The difference between these two
variants can easily be inferred from Fig.14.

\bigskip

{\bf Acknowledgments}. The authors want to thank the Deutsche
Forschungsgemeinschat for financial support. Two of us (F.A.I and
V.V.P.) would like to thank the Physics Department of the TUM for the
hospitality extended to them during their stay in Garching.

\appendix
\section{Intrinsic response function}

The computation of the intrinsic response function is somewhat involved.
Due to the frequency dependence both of Fermi distribution $n(\omega)$
and width $\Gamma(\omega)$ the integral in (\ref{respfun}) can not be
calculated analytically. The numerical integration is rather time
consuming since $\varrho_k(\omega)$ are sharply peaked functions which
width varies by two order of magnitude depending on the values of $e_k$.
Fortunately the integration in (\ref{respfun}) can be carried out by
means of residues theorem closing the integration limit in the lower
half plane. For this one needs to find the poles and residues of all the
terms in the integrand of (\ref{respfun}).

Let us first look for the poles of ${\cal G}_k(\omega +i\epsilon)$.
Substituting (\ref{dspspecdensgam}),(\ref{imselfenomt}) into
(\ref{green}) and introducing notation $\Delta \equiv
\sqrt{c^2+\pi^2T^2},~\mu^{\pm}=
\mu \pm i \Delta$ (\ref{green}) can be brought to the form
\begin{eqnarray} 
&{\cal G}_k(\omega +i\epsilon)= (\hbar \omega -\mu^+)
(\hbar \omega -\mu^-)\times \nonumber\\ 
&\Big\lbrace(\hbar \omega
-e_k)(\hbar \omega -\mu^+)(\hbar \omega -\mu^-) +{[\hbar \omega
-\mu^++i\Delta]\over 2\Gamma_0 \Delta /c^4} +{i[(\hbar \omega
-\mu^+)(\hbar \omega -\mu^-)-c^2]\over 2\Gamma_0 /c^2} \Big\rbrace^{-1}
\label{greenp}
\end{eqnarray}
It is not difficult to note that the last terms in square brackets
cancel with each other and both numerator and denominator can be divided
by $\hbar \omega - \mu^+$. In this way ${\cal G}_k(\omega +i\epsilon)$
becomes
\begin{equation}
\label{greenpp}
{\cal G}_k(\omega +i\epsilon)={{\hbar \omega -\mu^-}\over {(\hbar \omega
-\hbar \omega^+_k)(\hbar \omega -\hbar \omega^-_k)}} ={1\over
{\hbar^2(\omega^+_k-\omega^-_k)}} \left[{{\hbar \omega^+_k -\mu^-}\over
\omega -\omega^+_k} +{{\mu^-\hbar \omega^-_k}\over\omega
-\omega^-_k}\right]
\end{equation}
where $\hbar \omega_k^{\pm}$ are solutions of the equation
\begin{equation}
\label{zeros}
(\hbar \omega -\hbar \omega^+_k)(\hbar \omega -\hbar \omega^-_k)\equiv
\left(\hbar \omega -e_k +i{c^2\over 2\Gamma_0}\right)(\hbar \omega
-\mu^-)+ {c^4\over 2\Gamma_0 \Delta} = 0
\end{equation}
namely
\begin{equation}
\label{tpoles}
\hbar \omega^{\pm}_k={1\over 2}\Big\lbrace e_k +\mu^-
    -i{c^2\over 2\Gamma_0}\pm \left[\left(e_k -\mu^-
    -i{c^2\over 2\Gamma_0}\right)^2-{2c^4\over \Gamma_0 \Delta}
    \right]^{1/2} \Big\rbrace    
\end{equation}
The both poles of  ${\cal G}_k(\omega +i\epsilon)$ lie in the lower
half-plane.  The residues of ${\cal G}_k(\omega +i\epsilon)$ are simple
functions of $\hbar \omega_k^{\pm}$ and $\mu^-$ as it is seen from
(\ref{greenpp}).  From the definition (\ref{green}) it is easy to seen
that poles and residues of ${\cal G}_k(\omega -i\epsilon)$ are complex
conjugated to that of  ${\cal G}_k(\omega +i\epsilon)$.  The pole
representation for $\varrho_k(\omega)$ is easily obtained from
(\ref{greenpp}).  Besides the poles of ${\cal G}^{\pm}_k(\omega\pm
i\epsilon )$ and $\varrho_k(\omega)$ one should account also for the
poles of Fermi function $n(\omega)$ in the plane of complex  $\omega$
(so-called Matzubara frequencies)
\begin{equation}
\label{mazubara}
\hbar \omega_n = \mu \pm i\pi T (2n+1),\,\,\,\,\,\,n=0,1,2,... 
\end{equation}
In principle, the sum extends over infinite many terms, but
in practice the summation in (\ref{respfun}) is cut at such frequencies
$\omega_n$  which contribute less than determined by the desired
accuracy.

\vskip 1cm


\vfill

\end{document}